%% file: InterpAMR.tex
\documentclass[review]{elsarticle}

\input{PREAMBULE}

\begin{document}

\input{FRONTMATTER}

\section{Introduction}
\input{Intro}


\section{Grid specification} 
\input{Grid}

\section{Basic concepts and definitions}
\label{Basics}
\input{Basic_notation}

\section{Full algorithm description}

\subsection{General idea and continuity requirements}
\input{General_algorithm}

\subsection{Algorithm implementation}

\input{Implementation}





\section{Testing the algorithm}

\input{Test}

\section{Conclusion}

\input{Conclusion}

\section*{References}

\bibliography{mybibfile}

\end{document}

%% file: PREAMBULE.tex
\usepackage{amsmath}
\usepackage{graphicx}
\usepackage{lineno}
\usepackage{verbatim}
\usepackage{array}
\usepackage{subfigure}

\journal{Journal of \LaTeX\ Templates}

%% file: FRONTMATTER.tex
\begin{frontmatter}

\title{An Efficient Second-Order Accurate and Continuous Interpolation
  for Block-Adaptive Grids}


\author[mainaddress]{Dmitry Borovikov}

\author[mainaddress]{Igor V. Sokolov}

\author[mainaddress]{G\'abor T\'oth}

\address[mainaddress]{Department of Atmospheric, Oceanic and Space Science, 
University of Michigan, 2455 Hayward St., Ann Arbor, MI 48109-2143}

\begin{abstract}
  \input{Abstract}

\end{abstract}

\end{frontmatter}

%% file: Abstract.tex
In this paper we present
a second-order and continuous interpolation algorithm
for cell-centered adaptive-mesh-refinement (AMR) grids.
Continuity requirement poses a non-trivial problem 
at resolution changes. 
We develop a classification of the resolution changes,
which allows us to employ efficient and simple 
linear interpolation in the majority of the computational domain. 
The benefit of such approach is higher efficiency.
The algorithm is well suited for massively parallel computations.
Our interpolation algorithm allows extracting
jump-free interpolated data distribution
along lines and surfaces within the computational domain.
This capability is important for various applications,
including kinetic particles tracking 
in three dimensional 
vector fields, 
visualization (i.e. surface extraction)
and extracting variables along one-dimensional curves such as field lines,
streamlines and satellite trajectories, 
etc.
Particular examples of the latter are
models for acceleration of solar energetic particles (SEPs) 
along magnetic field-lines. 
As such models are sensitive to sharp gradients and discontinuities
the capability to interpolate the data from the AMR grid
to be passed to the SEP model without producing false gradients numerically
becomes crucial. 
The code implementation of our algorithm is publicly available
as a Fortran 90 library.

%% file: Intro.tex
The adaptive-mesh-refinement (AMR) grids \cite{Berger1989JCP}
have become an essential part 
of many applications in computational physics.
The AMR is an effective technique which allows us to adapt grid to particular features being simulated.
For example adaptive spatial discretization is employed by 
the Block-Adaptive Tree Solarwind Roe-type Upwind Scheme (BATS-R-US) 
code for solving MHD equations \cite{Powell1999,Powell2003,Gombosi2004CompSEng,Powell2005}, 
which is the heart of Space Weather Modeling Framework (SWMF)  \cite{Toth2005JGR, Toth2007, Toth2012}.

Often the physical quantities to be calculated are 
sampled at cell centers especially within the framework of
control volume method. 
To obtain data at different locations one needs an interpolation algorithm.
A high variety of approaches allows us to achieve the second order of accuracy.
While for many applications this is enough to obtain accurate results, 
certain applications  are sensitive to continuity, and therefore
require an interpolation method that does not 
generate artificial discontinuities.
For such cases an algorithm is needed, which 
meets requirements of both continuity and second order of accuracy. 

A particular application that motivated us to develop such algorithm is acceleration
of solar energetic particles (SEPs) at the shock wave fronts.
The model presented in \cite{Sokolov2004APJ, Sokolov2006APJ,Sokolov2008AIPC, Sokolov2009APJ} 
employs magnetic field line tracing
as well as interpolation of solar wind parameters 
from the BATS-R-US AMR grid to 
the locations of the magnetic field lines points.
Unphysical discontinuities within the interpolation algorithm affect results 
by producing particle acceleration/deceleration 
at false shocks near grid irregularities.
Hence, the model's reliability is crucially dependent 
on the continuity of interpolation
algorithm.

The problem of AMR interpolation has a variety of possible applications,
under which it has been addressed before 
\cite{Colella2007JPCS,Borovikov2006ASPC,MacNeice2000,Sokolov2006JCP}.
In the latter reference, the authors considered it in the context of 
the total-variation-diminishing (TVD) principle 
applied to block-adaptive grids.
Interpolation is an important part of the limited reconstruction procedure,
the choice of the interpolation algorithm affects 
stability and accuracy of a scheme
it is being applied to. 
Another possible application is  visualization of the result 
obtained in simulations with AMR grids.
In this context
the problem of continuous AMR interpolation has also been addressed
\cite{Weber2001DataVis,Weber2001VMV,Moran2011IEEE}.
Therefore, development of such algorithm is not 
a problem of merely mathematical interest 
but it is also of importance in applications.

The numerical results obtained by control volume method in conjunction 
with AMR are sampled on the cell-centered grid.
To visualize data, one way is to resample them 
to corner-centered grid as done by many 
visualization tools, such as Tecplot.
However, this approach involves more distant grid cells
and, as a result, smoothens the data.
Alternative way is to visualize the uniform parts of the AMR grid directly
and to create a smooth transition between 
parts of the grid at different resolution levels.
To do so in \cite{Weber2001DataVis,Weber2001VMV,Moran2011IEEE},
the authors developed a stitching algorithm, 
as described in details in \cite{Weber2001DataVis}.

Their method achieves continuity and the second order of accuracy.
Tessellation of the computational domain into simpler shapes is employed, 
and the interpolation procedure is fully defined 
by the geometry of a particular shape
a given point falls into.
The algorithm is easy to understand and, hence, simple to implement.
However, when developing our algorithm, we focused on its efficiency.
Compared to the stitching algorithm, it is advantageous
in the sense that it uses the same simple interpolation procedure
over large portion of the computational domain.
As described further, the stitching algorithm is 
double-branched in two dimensions (2D)
and triple-branched in 3D, 
while ours stays uniform.
Therefore the latter is preferable for 
repeated interpolation during the simulation.

%% file: Grid.tex
Here, we focus on block-adaptive AMR grids and assume that 
the computational domain is decomposed into blocks, 
each block being a rectangular box in Cartesian (or logically Cartesian)
coordinates.  The blocks are decomposed into 
$i\times j\times k$ ``cells'', 
for three-dimensional (3D) case, the cell-per-block integer numbers 
(usually, even) being constant throughout the whole grid, 
which maintains the claimed blocks similarity. Being similar, the blocks, however, are not all 
identical, as long as the cell size in different blocks may differ. 
Specifically, we assume that in the neighboring blocks 
(having at least one common point on the boundary) 
the cell size, $\Delta x^{(C)},\,\Delta y^{(C)},\,\Delta z^{(C)}$, 
in the {\it coarser} (C) block may be 
by a factor of two larger than those in the  {\it finer} (F) block: 
\begin{center}
        $\Delta x^{(C)}=2\Delta x^{(F)},\,\Delta y^{(C)}=2\Delta y^{(F)},\,\Delta z^{(C)}=2\Delta z^{(F)}$.
\end{center} 
Refinement ratios more than 2 are also possible 
and the algorithm generalizes to these cases, 
but our implementation is restricted to the refinement ratio of 2.


We assume that
the numerical solution of the governing equations, 
obtained at each time step, is a {\it cell-centered} grid function, 
e.g. the solution obtained using {\it control volume} method.
In order to find a numerical solution at an 
arbitrary point within the computational domain one needs to find the way 
to interpolate data from the cell-centered block-adaptive AMR grid. 
The goal of the present work is to find the procedure to interpolate data 
from cell-centered block-adaptive logically Cartesian AMR grid, 
which continuously connects bilinear interpolation from 
the uniform parts of the grid through the resolution changes. 

Indeed,  
for a uniform 2D Cartesian grid the easiest and most natural approach is 
a bilinear interpolation (for three dimensions it is 
trilinear interpolation). It is efficient and of the second order accuracy.
Another advantage is that type of interpolation retains symmetries with respect 
to the coordinate axes if they are assumed by a correspondent symmetry of the 
problem to be solved. Should one perform the interpolation by splitting 
the computational domain into a set of tetrahedra, 
the second order accuracy would be also achieved, however, the symmetry 
of the numerical solution will be broken, as long as the set of tetrahedra 
would not be symmetric. In addition, this partitioning is not a unique solution, 
thus making ambiguous the interpolation procedure. The latter may be unacceptable.

Compared to tetrahedron-based method trilinear interpolation has one important
drawback: it is not continuous when directly applied 
near grid resolution changes.
This requires us to generalize the trilinear 
interpolation method for these problematic sites, as we describe below.

%% file: Basic_notation.tex
\label{sec:basic}
In this section we describe basic definitions and methodology. Many ideas naturally translate
from lower to higher dimensions. 
For this reason notations have been developed in an arbitrary number of dimensions $N$.

For any arbitrary point {\bf X} an algorithm solves two distinct problems:
to determine an {\it interpolation stencil} (set of cells involved into interpolation
with non-zero weights) and to calculate {\it interpolation weights}.
An interpolation stencil should consist of cells that are close to {\bf X}.
In order to elucidate ``closeness'' we introduce the following notions.

\textit{Enclosing stencil} for point {\bf X} 
(the point {\bf X} is enclosed by this stencil) 
is a set of $2^N$ grid cell centers
with their coordinate vectors, 
$\mathbf{x}_1$, ... , $\mathbf{x}_{2^N}$, 
satisfying the following two conditions:

First, the rectangular box bounded by their coordinates
contains point $\mathbf{X}=(X,Y,Z)$.
For example, for 2D grids cell centers in the enclosing stencils 
for point {\bf X} must satisfy inequalities:
\begin{linenomath}\begin{align}
        x_1 \le X < x_2,&  \hspace{20pt} y_1 \le Y < y_3, \nonumber\\
        x_3 \le X < x_4,&  \hspace{20pt} y_2 \le Y < y_4. 
        \label{eq:Enclosing1}
\end{align}\end{linenomath}
In 3D case for the first four points, 
$\mathbf{x}_1$, .., $\mathbf{x}_4$,  
the inequalities \eqref{eq:Enclosing1} are fulfilled together with
the requirement, $z_i \le Z$, where $i=1,..,4$.
For the last four points, $\mathbf{x}_5$, .., $\mathbf{x}_8$,  
the inequalities analogous to \eqref{eq:Enclosing1} are fulfilled 
together with the requirement, $Z < z_i$, where $i=5,..,8$.

Second, each edge of that box does not exceed 
the linear size of Coarser cells, $\Delta x^{(C)}$, $\Delta y^{(C)}$, 
along the corresponding axis, $i$, e.g. for 2D grids:
\begin{linenomath}\begin{align}
        &|x_{1,3} - x_{2,4}| \le \Delta x^{(C)}, \nonumber\\
        &|y_{1,2} - y_{3,4}| \le \Delta y^{(C)}.
        \label{eq:Enclosing2}
\end{align}\end{linenomath}
As a part of definition, we introduce notion of {\it Fine cluster}.
Two Finer grid points of the enclosing stencil form a {\it 2-cluster},
if they lay on a line parallel to a coordinate axis,
while four Finer grid points form a {\it 4-cluster}, if they lay on a plane
parallel to a coordinate plane, etc. 
To reduce an ambiguity, we set an extra constraint on an enclosing stencil,
by claiming that the set of conditions \eqref{eq:Enclosing2}
should be strict inequality for stencil points, e.g. 1 and 2,
if both belong to the same Fine cluster, 
e.g. $|x_1 -  x_2| < \Delta x^{(C)}$.
In this way we exclude from consideration the enclosing stencils, 
which unreasonably involve farther Finer grid points instead of
closer ones.

\begin{figure}[!h]
  \centering
  \includegraphics[width=4.0cm]{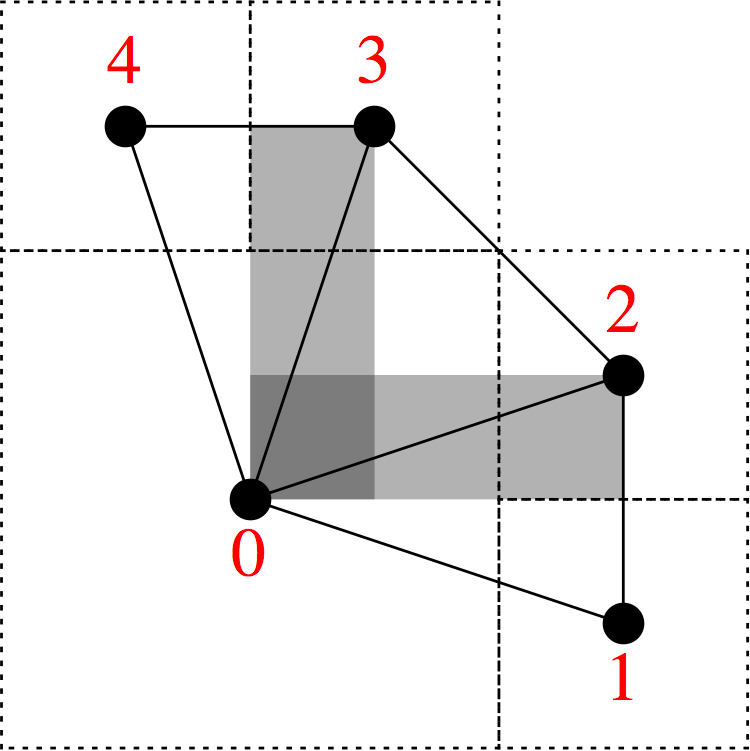}
  \caption{Ambiguity of an enclosing stencil. All points inside the shaded rectangles
           are enclosed by 2 different stencils: 
           (0,1,4,2) and (0,1,4,3) for dark gray,
           (0,1,3,2) and (0,1,4,2) for horizontal light gray, 
           (0,1,4,3) and (0,2,4,3) for vertical light gray.}
  \label{fig:2DAmbiguity}
\end{figure}

Still, we emphasize that an enclosing stencil is not unique for many point locations,
as illustrated in Figure \ref{fig:2DAmbiguity}.
For this reason it can't be identified with an interpolation stencil,
which must be unambiguous.
However, all stencils enclosing the same point have 
the same values of {\it edge type}.
Herewith, 
we say that a stencil has an edge type $n$, $n\le N$, 
if it has resolution changes in $n$ dimensions.
Within this approach, $(N{-}n)$-dimensional rectangles, 
which are shaped by points of a stencil parallel to the remaining dimensions,
are refered to as {\it trivial elements} of this stencil.
Interpolation procedure on them is a simple generalization
of a linear interpolation.

We refer to a set of points satisfying \eqref{eq:Enclosing1} 
for a given enclosing stencil
as \textit{an enclosed set} of this stencil.
An enclosed set is an $N$-dimensional rectangular box.
Adjacent enclosed sets of 
enclosing stencils of the same edge type, $n$,
merge together into a \textit{resolution n-edge}.
Particularly, 0-edge (i.e. no resolution change) 
is a closed isolated domain covered by a uniform Cartesian grid.
The following easy-to-prove claims are important.
First, any given point can belong to one and only one resolution edge.
In other words, the computational domain decomposes into a set of  
resolution $n$-edges.
Second, for $n<N$, 
one can use simple linear interpolation in all dimensions that do not 
have a resolution change, i.e. on trivial elements.
Thus, the effective dimensionality of non-trivial interpolation procedure
for $n$-edge reduces to $n$.
Indeed, 
this non-trivial interpolation procedure should be introduced 
on an $n$-dimensional subspace 
orthogonal to trivial elements. 

Therefore, we define a \textit{main interpolation subset} 
for point {\bf X} inside a resolution $n$-edge  
as a cross-section of this $n$-edge by $n$-dimensional plane 
that includes point {\bf X} and 
is perpendicular to trivial elements of enclosing stencils 
for the point {\bf X}. 
Particularly, for $n{=}N$ main interpolation subset coincides 
with the resolution $n$-edge itself.
For 1-edge we refer to it as {\it main interpolation line}
(see Figure \ref{fig:1edge}), 
for 2-edge - {\it main interpolation plane} (see Figure \ref{fig:2edge}).

As the last definition, an {\it extended interpolation stencil} 
for a point {\bf X} is a minimal union of 
stencils enclosing  together the main interpolation subset for this point. 
Unlike an enclosing stencil, the extended interpolation stencil 
is unique for any given point.

This gives us a starting point to outline the following algorithm.
First, for a given point {\bf X} we need to figure out the type of $n$-edge
it belongs to.
Based on found value of $n$ and point location, one needs: 
(1) to construct the extended interpolation stencil;
(2) to choose the actual interpolation stencil from the extended interpolation stencil;
and
(3) to calculate the interpolation weights. 
Then, with grid points forming the interpolation stencil for {\bf X}, 
indexed by $i=1, ..,m$, 
interpolation weights, $w_i$,
and values of a function $f$ sampled at grid points, $f_i$, 
the interpolated value in {\bf X} is calculated as 
$f(\mathbf{X})=\sum_{i=1}^m{w_i f_i}$.
We note that our algorithm does not use ghost cells. 
In a parallel implementation, the sum can be calculated with
an \texttt{MPI\_reduce} call, for example.

\begin{figure}
 \centering
    \begin{tabular}{cc}
      \includegraphics[width=4cm]{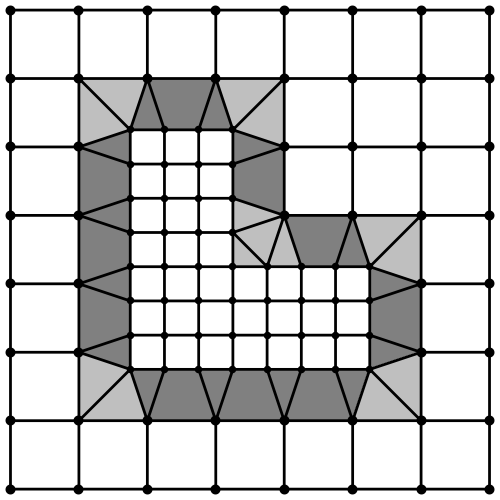}
      \includegraphics[width=4cm]{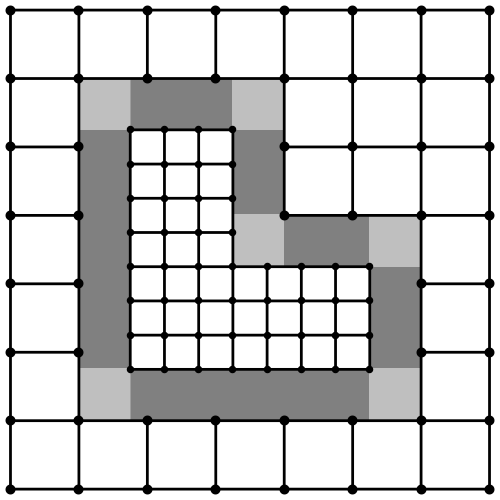}
    \end{tabular}
    \caption{Stitching (see \cite{Weber2001DataVis}) and 
             resolution edges 
             (white - 0-edges, dark gray - 1-edges, 
             light gray - 2-edges) on 2D grid
             of $4\times 4$ adaptive blocks. 
             Vertices are cell centers of the actual grid.}
    \label{fig:2DCompare}
\end{figure}

To conclude this section, we compare stitching algorithm \cite{Weber2001DataVis}
versus splitting the domain into $n$-edges. 
The left panel of Figure \ref{fig:2DCompare} is based on \mbox{Figure 3} 
of \cite{Weber2001DataVis} 
showing a 2D AMR grid with stitch cells along edges 
of resolution regions (dark gray) 
and near their corners (light gray).
On the right panel one can see 1-edges (dark gray) and 2-edges (light gray) on the same grid.
Though the images look similar, the boundaries between 
the light and dark gray zones are somewhat different. 
The stitching algorithm is easier in description
as it decomposes resolution change regions into
simple shapes and performs interpolation based 
within them. 
However, throughout the 1-edge zone
we benefit from applying a uniform algorithm as described below, 
while for the stitching algorithm
two different sorts of stencils are employed.
In 3D case the stitching algorithm branches further,
employing three sorts of stencils for regions
corresponding to a resolution 1-edge.
As computational time on resolution changes
is dominated by resolution 1-edges,
our algorithm is more efficient.

%% file: General_algorithm.tex
Our goal is to develop a consistent algorithm that generalizes 
bilinear and trilinear interpolation to block-adaptive grids.
For this reason, we apply bilienar/trilinear interpolation 
on uniform parts of the grids, which are 0-edges,
as well as on trivial elements of extended interpolation stencils
for higher order edges.

For resolution edges of types $n=1, .., N$ on $N$-dimensional grid, 
our interpolation algorithm employs values
in intersection points of the main interpolation subset 
with trivial elements of extended interpolation stencils. 
These values are calculated using linear/bilinear interpolation on
$(N{-}n)$-dimensional trivial elements of extended interpolation stencils.
Then, the interpolation scheme on resolution $n$-edge is solved,
independently of the actual dimensionality of the grid, $N$,
by applying the calculated interpolation weights to the values interpolated
to the intersection points.
Hence, the resulting interpolation weights are products of weights used
to obtain the values at the intersection points and the weights
resulting from interpolation procedure on a resolution $n$-edge.

The requirements of continuity yields an obvious  
relation between different types of edges:
on interfaces between resolution $n$-edges the interpolation scheme
should reduce to the algorithm used in adjacent
edges of lower edge type. 


%% file: Implementation.tex
The code implementation of the algorithm described in this section
is publicly available as a Fortran 90 library,
\texttt{ModInterpolateAMR.f90},
currently as a part of the SWMF distribution.

As outlined above, the interpolation algorithm for a given point {\bf X}
starts from determining the type of edge it belongs to
and constructing the extended interpolation stencil.
Practically, these two steps are combined together in the following manner.
We assume in our presentation that the whole block-adaptive grid
is described in terms of a ``find'' procedure, 
which for any given point coordinates returns 
the indices of the grid block and the grid cell containing the point
as well as the cell sizes 
$\Delta x$, $\Delta y$, $\Delta z$, 
of the block 
and the point coordinates with respect to the block corner.

\subsubsection{Construction of an extended stencil}

Now, for the point {\bf X} within the computational domain 
we first ``find'' the {\it initial} block it falls into.
We note that we apply the proper nesting \cite{Berger1989JCP}
restriction on the refinement levels so that grid levels of
adjacent blocks (including diagonal directions) cannot 
differ by more than one.

If {\bf X} happens to lay in the block interior,
specifically farther than half a cell size apart from any block boundary,
then it is necessarily within a uniform part 
of the grid formed by the block cell centers.
Therefore, in order to improve the time performance,
in this case a bilinear/trilinear interpolation 
is applied immediately
and algorithm quits, returning weights and indices 
for the grid points of the enclosing
stencil (a rectangular box 
$\Delta x^{(C)}\times\Delta y^{(C)}\times\Delta z^{(C)}$, the cell sizes
for the {\it initial} block, they are marked as Coarse for the reason explained below) 
as the final interpolation stencil.
In 2D case the weights of a bilinear interpolation, $w^{2D}_i$, $i=1, ..,4$,
are calculated in terms of the components $(C_x, C_y)$ of the dimensionless
coordinate  vector of a given point {\bf X} with respect to the first vertex 
of the enclosing stencil: $C_x=(X-x_1)/\Delta x^{(C)}$, $C_y=(Y-y_1)/\Delta y^{(C)}$.
The interpolation weights of a 2D bilinear interpolation are:
\begin{linenomath}\begin{align}
        w^{2D}_1&= (1-C_x)(1-C_y)      &w^{2D}_2&= (1-C_x) C_y \nonumber\\
        w^{2D}_3&=    C_x (1-C_y)      &w^{2D}_4&=    C_x  C_y.
        \label{eq:2linear}
\end{align}\end{linenomath}
In 3D yet another dimensionless coordinate is used, 
$C_z=(Z-z_1)/\Delta z^{(C)}$,
and the trilinear interpolation weights are:
\begin{linenomath}\begin{align}
        w^{3D}_i&=w^{2D}_i (1-C_z),\quad i=1, ..,4, \nonumber\\
        w^{3D}_i&=w^{2D}_{i-4} C_z,  \quad i=5, ..,8.
        \label{eq:3linear}
\end{align}\end{linenomath}

Otherwise, we extend the cell-centered grid beyond the block boundary.
Now point {\bf X} is enclosed by some rectangular box $\Delta x^{(C)}\times \Delta y^{(C)}\times \Delta z^{(C)}$
of the extended grid, but only some vertices of the box lay within 
the {\it initial} block. 
For the other vertices we ``find'' the block(s) those vertices fall into.
If the newly ``found'' block(s) are at the same resolution level
as the {\it initial} block,
the vertices of the enclosing stencil coincide with the cell centers
in these block(s) so that the indices for these blocks and cells
should be included into the interpolation stencil.
As above, the bilinear/trilinear interpolation is applied 
and the algorithm quits.

Otherwise, if any of the newly ``found'' block(s) is at the Finer resolution level,
we need to form an extended stencil for {\bf X} based on the coordinates
of vertices of a rectangular box of size 
$\Delta x^{(C)}\times\Delta y^{(C)}\times\Delta z^{(C)}$ that we call the 
{\it Coarse-cell sized box} (CSB).
An extended stencil includes grid points (their coordinates, cell and block ids)
of the Coarser blocks, which coincide with the vertices of the CSB,
as well as the Fine $2^N$-clusters in the Finer blocks,
the center of the cluster coinciding with the vertices of the CSB.
Otherwise, if any of the newly ``found'' blocks is Coarser than 
the {\it initial} one,
we claim this Coarser block to be {\it initial} and restart constructing
the CSB and the extended stencil.

It is easy to see that:
(1) CSB is unique for any point within the computational domain;
(2) point {\bf X} belongs to the CSB, 
    but doesn't belong to any Fine $2^N$-cluster,
    therefore, herewith, by CSB we mean the CSB 
    excluding the domains enclosed by its Fine $2^N$-clusters;
(3) hence, the procedure above decomposes the computational domain
    into CSBs and rectangular boxes;
(4) union of all rectangular boxes is a union of all resolution 0-edges,
    while union of all CSBs is a union of all resolution edges of higher edge type;
(5) the constructed set of grid points includes redundant points,
    but is guaranteed to include the extended stencil for 
    any point {\bf X} within the CSB.

Upon this stage of the algorithm,
for a given point {\bf X}
we either have found it belonging 
to a resolution 0-edges and performed a bilinear/trilinear interpolation, 
or have constructed a set of grid points, 
which includes the extended interpolation stencil for {\bf X}.

\subsubsection{Solving for edge type}

The input for this stage of the algorithm is a set of grid points
found in the previous stage and resolution levels of CSB vertices
({\it level sequences}).
The expected output is an interpolation stencil, 
which is 
chosen from grid points of this set.
It is easy to construct an enclosing stencil for the CSB {\it central} point,
by taking all Coarser vertices of the set and selecting a single grid point
from each Fine $2^N$-cluster,
which is the closest to the central point.
Herewith, we refer to this stencil and the body it shapes 
as {\it central shape} in general 
or either {\it central quadrangle} in 2D, or {\it central hexahedron} in 3D.
The central shape determines the upper limit for the edge type
of the resolution edges present in the decomposition of the CSB,
which can be derived from the sequence of the refinement levels.
For example, the level sequence (the order of vertices 
in all sequences is defined as in Equation \eqref{eq:Enclosing1})
(CCCCFFFF) determines a resolution 1-edge 
(interface perpendicular to $z$-axis, 
see the right panel in Figure \ref{fig:1edge}),
the level sequence (CFFFCFFF) determines a resolution 2-edge
(edge going along $z$-axis, 
see the bottom left panel in Figure \ref{fig:2edge}).
As we see, the level sequence of CSB determines 
not only the edge type, but also direction of trivial
elements (if present) of a central shape
and its orientation.
We construct a lookup table for 
$2^{2^N}{-}2$ possible level sequences of CSB (excluding the 2 uniform cases), 
which allows us to efficiently determine the
edge type and to identify the particular configuration. 

\subsubsection{CSB decomposition and interpolation: resolution 1-edge}
\label{sec:1edge}

\input{CSB.1edge}

\subsubsection{CSB decomposition and interpolation: resolution 2-edge}
\label{sec:2edge}

\input{CSB.2edge}

\subsubsection{CSB decomposition and interpolation: resolution 3-edge}
\label{sec:3edge}

\input{CSB.3edge}

\subsubsection{Interpolation on simple shapes}

\input{simpleshapes}

%% file: CSB.1edge.tex
\input{fig.1edge}

The simplest case is if the central part of the CSB, 
hence, the whole CSB, is a resolution 1-edge.
In this case 
the interpolation stencil for {\bf X},
which coincides with its enclosing stencil,
consists of Coarser vertices  and the Fine $2^{N{-}1}$-cluster, 
which {\bf X} projects onto.

The interpolation procedure is a linear interpolation
applied to endpoints of the main interpolation line,
which are 2 vertices in 1D case or intersection points
of the main interpolation line with trivial elements of enclosing stencil
in 2D and 3D,
the endpoint values are obtained using linear/bilinear interpolation 
on trivial elements as shown in Figure \ref{fig:1edge}.
As mentioned before, 
we apply this simple algorithm throughout 
the whole resolution 1-edge. 
The algorithm quits.

This approach is different from that developed in \cite{Weber2001DataVis},
where the authors decompose corresponding part of 
a grid into two sorts of shapes
in 2D and three sorts of shapes in 3D based on position of the point {\bf X}
{\it within} this region.
This approach is easy to understand and implement,
but our algorithm is simpler and more efficient.
We apply the same simple interpolation procedure every time
point {\bf X} falls into a resolution 1-edge independently
of its position.

%% file: fig.1edge.tex
ß\begin{figure}[!h]
  \centering
    \begin{tabular}{cc}
      \includegraphics[height=4cm]{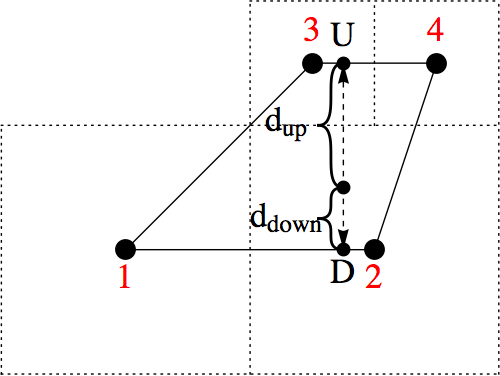}
      \includegraphics[height=4cm]{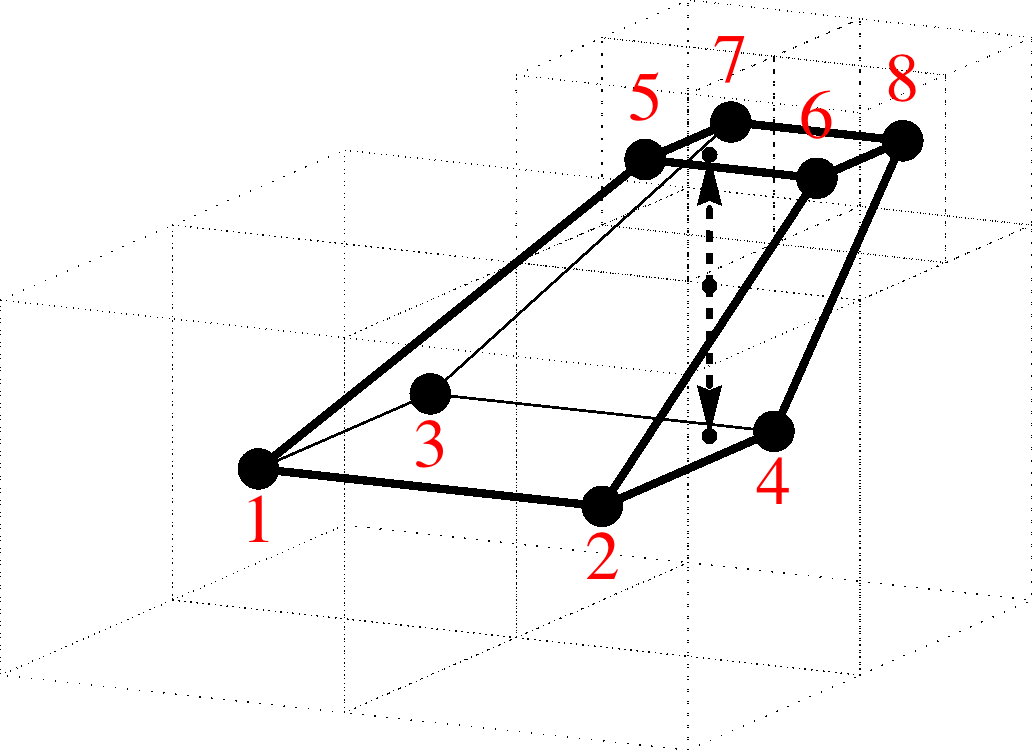}
    \end{tabular}
    \caption{Interpolation on 1-edge for 2D and 3D. 
             Interpolation weights are calculated using distances, 
             $d_{up}$ and $d_{down}$, 
             as: 
             $w_{up}=d_{down}/\left(d_{down}+d_{up}\right)$, 
             $w_{down} = 1 - w_{up}$,
             and are applied to the values in endpoints, $U$ and $D$.
             These values are calculated using linear as on the left
             and bilinear interpolation (see equation \eqref{eq:2linear}) 
             on the trivial elements as on the right panel,
             the latter yields weights $w_i^{2D}$, $i=1, ..,4$, 
             for lower face 
             and $w_i^{2D}$, $i=5, ..,8$, for upper face, .
             Then the final interpolation weights are
             $w_i=w^{2D}_i w_{down}$, $i=1,..,4$, 
             and $w_i=w^{2D}_i w_{up}$, $i=5,..,8$. 
             The continuity, for example, 
             at the boundary (1,2,3,4) between this 1-edge and 0-edge
             (see right panel)
             is ensured as our interpolation algorithm reduces 
             to bilinear interpolation when point {\bf X}
             approaches this boundary, both from below and from above.
             Also, as the main interpolation line crosses (2-4) edge,
             the weights of points 1 and 3 become zero,
             which ensures continuity within resolution 1-edge.
    }            
    \label{fig:1edge}
\end{figure}

%% file: CSB.2edge.tex
\input{fig.2edge}

If the edge type of the central shape derived from level set
appears to be 2,
then, generally speaking, the CSB decomposes into
resolution 1- and 2-edges.
Below we provide implementation details for search 
for the interpolation stencil.
Although not difficult for a resolution 2-edge,
decomposition of a resolution 3-edge becomes very sophisticated.
In order to reduce the computational time spent for the search,
we split CSB into rectangular boxes. 
Based on which particular box the point {\bf X} falls into,
we can eliminate some of the grid points from the extended stencil 
and reduce the number of simple shapes that may contain point {\bf X}.
We describe this procedure in details for 2-edge
and only briefly we outline this for 3-edge in the next section.

Note, that the effective dimensionality in this case is 2.
We consider a  2-edge in 2D ($x,y$) space,
or a 2-edge in 3D with trivial elements along $z$-axis. 
Let $x{=}x_{min}$ be the lower face (edge) 
of the CSB perpendicular to $x$-axis,
with $x{=}x_{min}+\Delta x^{(C)}{=}x_{max}$ being its upper face (edge).
We divide the CSB by a plane (line), 
$x{=}x_{min}+\Delta x^{(C)}/4$, 
if the face $x{=}x_{min}$ intersects at least one Fine cluster.
Analogously, we divide the CSB by a plane (line), 
$x{=}x_{max}-\Delta x^{(C)}/4$, 
if the face $x{=}x_{max}$ intersects at least one Fine cluster.
Repeating the same procedure for $y$-axis, 
we split CSB into rectangular boxes 
(see panels \subref{fig:2edge:2D1C}-\subref{fig:2edge:2D3C} 
in Figure \ref{fig:2edge}, splitting is shown in red).

The white rectangular domains in panel \subref{fig:2edge:2D3C} 
in Figure \ref{fig:2edge} are resolution 1-edges.
They are interpolated as described in section \ref{sec:1edge}
and the algorithm quits.

The shaded boxes form the main interpolation plane
for the point {\bf X} within the resolution 2-edge.
It is decomposed into triangles and we apply triangular interpolation.
Once we determine, which rectangular box 
contains {\bf X}, only a subset of triangles needs to be checked,
which makes the algorithm more efficient. 

Thus, the interpolation procedure within resolution 2-edge
is a triangular interpolation performed on selected vertices in 2D and
on selected intersection points of the main interpolation plane
with trivial elements in 3D.
Again, values in these intersection points are calculated using 
a linear interpolation on the trivial elements.
The product of weights used to calculate these values 
and those resulting from interpolation procedure 
in the main interpolation plane
yields final interpolation weights.
The algorithm quits.

We note that within resolution 2-edges of a 2D grid
our algorithm uses the same triangulation as \cite{Weber2001DataVis},
but it is connected to resolution 1-edges differently.
We also describe how CSB splitting into rectangular boxes
can be used to speed up the search for the interpolation stencil
(triangular stencil in this case).
Our approach for resolution 2-edges of a 3D grid
is different from \cite{Weber2001DataVis}, 
as the authors use a 3D tessellation, 
while we use a linear interpolation along the trivial elements
and a 2D interpolation on the main interpolation plane, 
which makes our algorithm simpler.


%% file: fig.2edge.tex
\begin{figure}[!h]
  \centering
  \renewcommand{\thesubfigure}{\Alph{subfigure}}
  \makebox[\linewidth][c]{
  \begin{tabular}{ccc}
    \subfigure[][]{
      \label{fig:2edge:2D1C}
      \includegraphics[width=3.5cm]{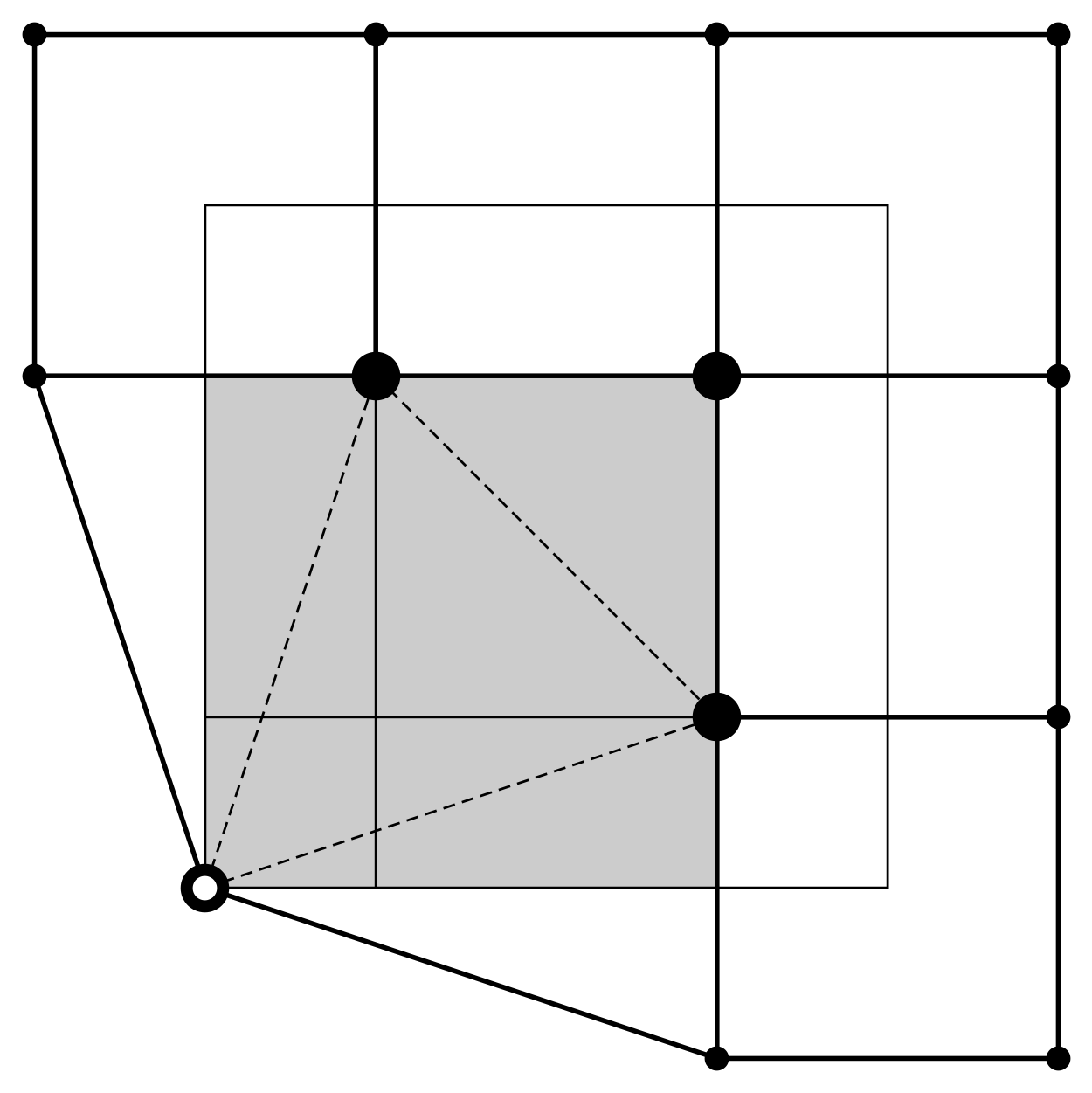} 
      }
    &
    \subfigure[][]{
      \label{fig:2edge:2D2C}
      \includegraphics[width=3.5cm]{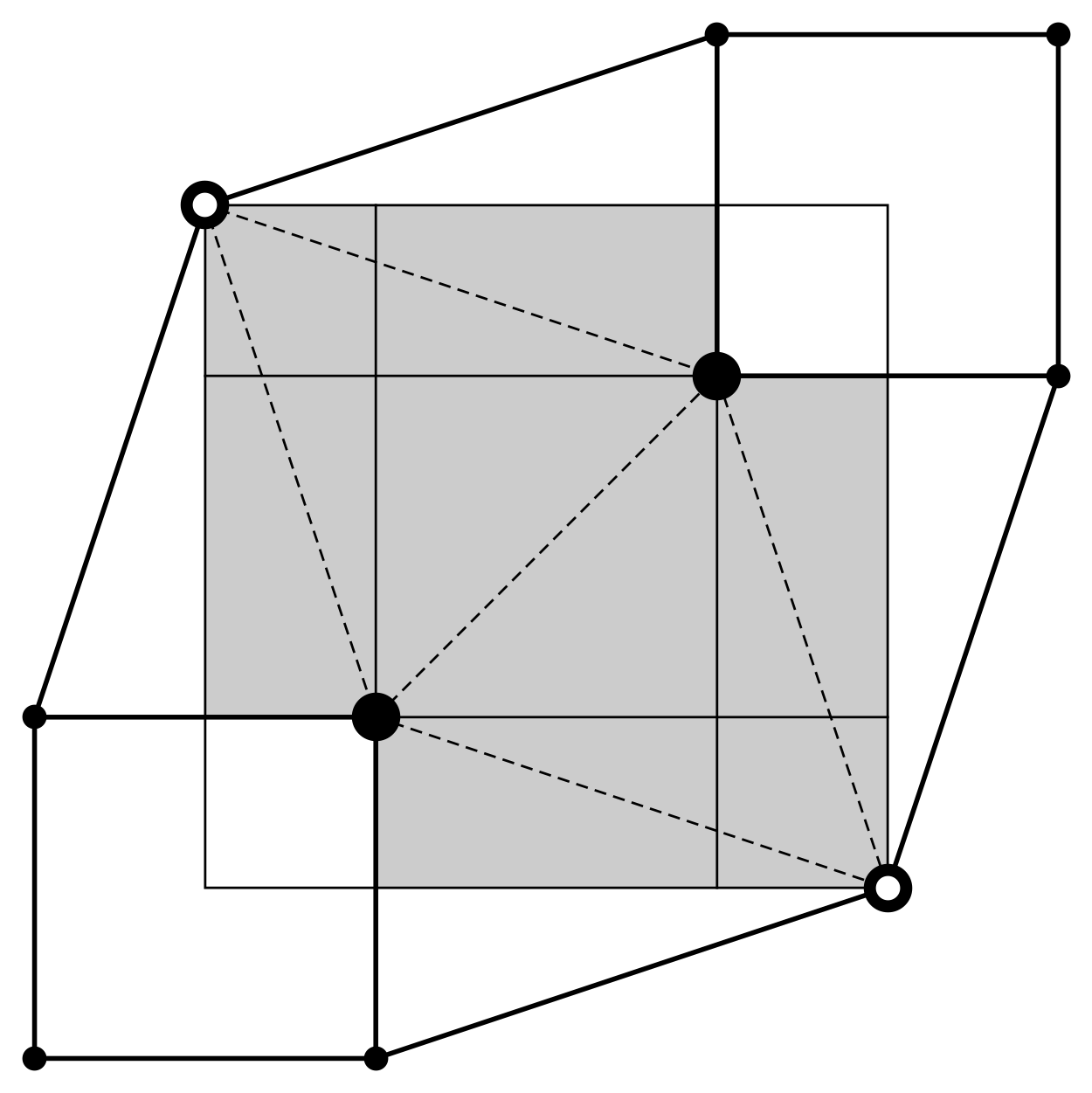} 
      }
    &
    \subfigure[][]{
      \label{fig:2edge:2D3C}
      \includegraphics[width=3.5cm]{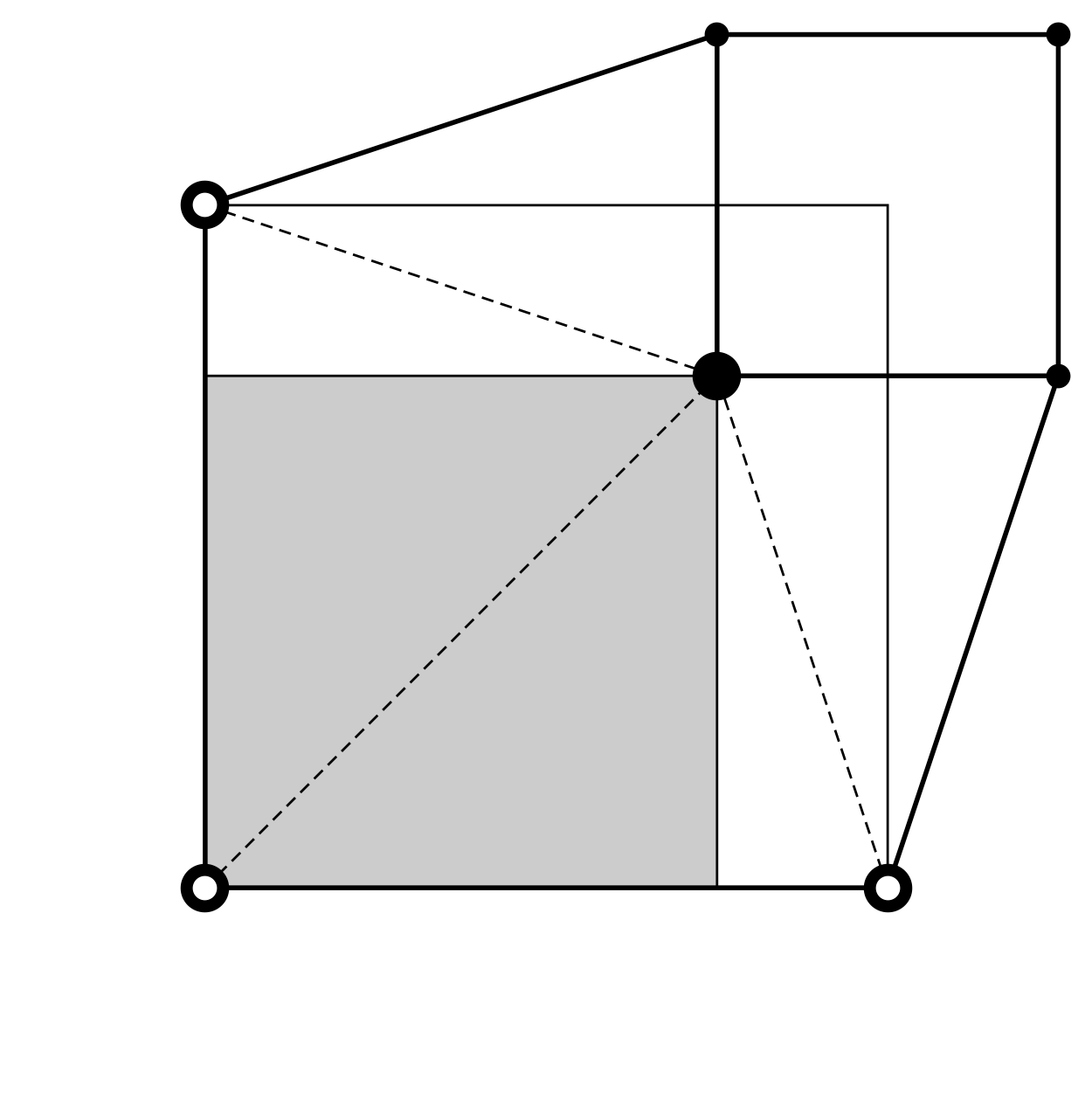} 
      }
  \end{tabular}
  }
  \vspace{-10pt}

  \makebox[\linewidth][c]{
  \begin{tabular}{cccc}
    \subfigure[][]{
      \label{fig:2edge:3D2C1}
      \includegraphics[height=3.5cm]{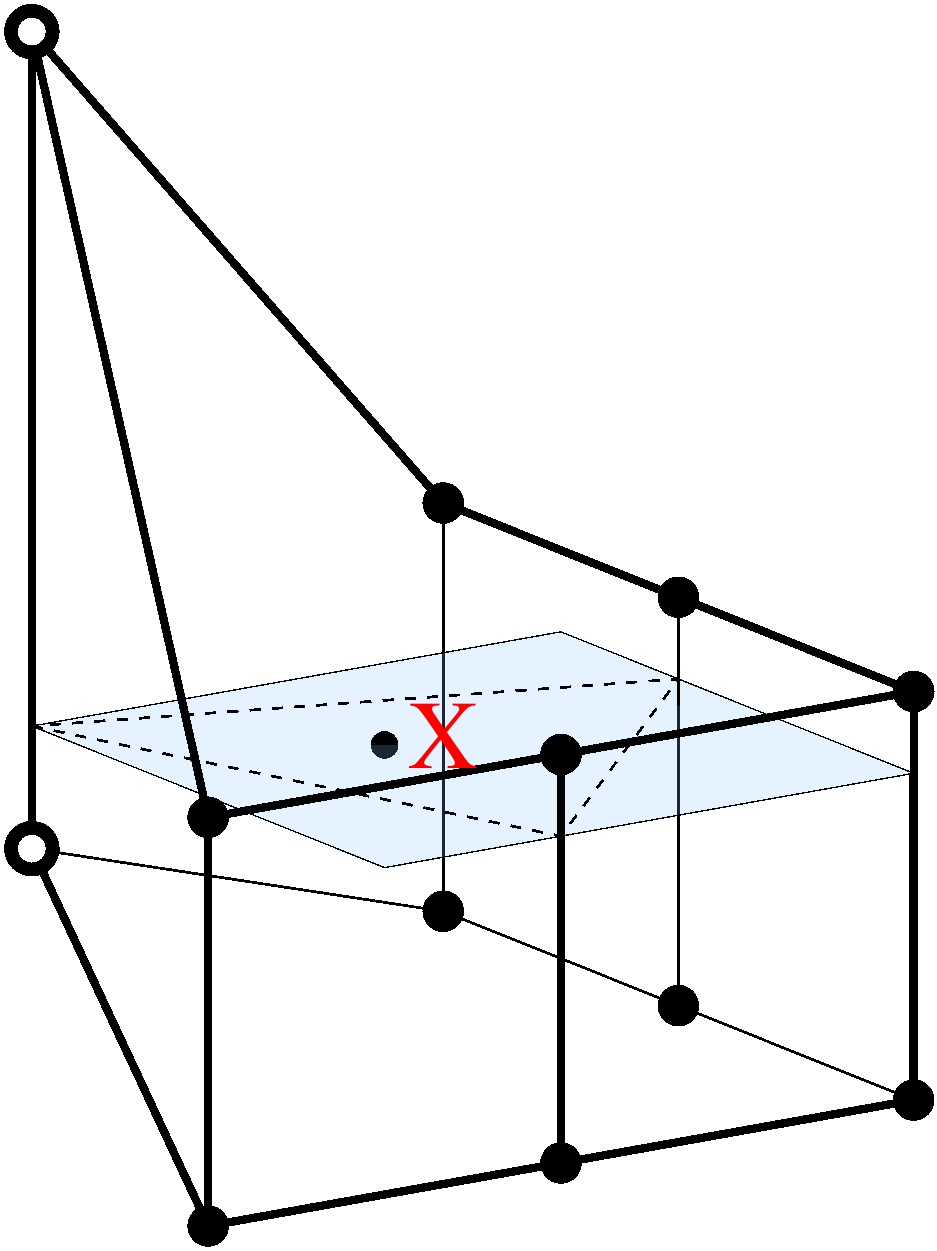} 
      }
    &
    \subfigure[][]{
      \label{fig:2edge:3D4C}
      \includegraphics[height=3.5cm]{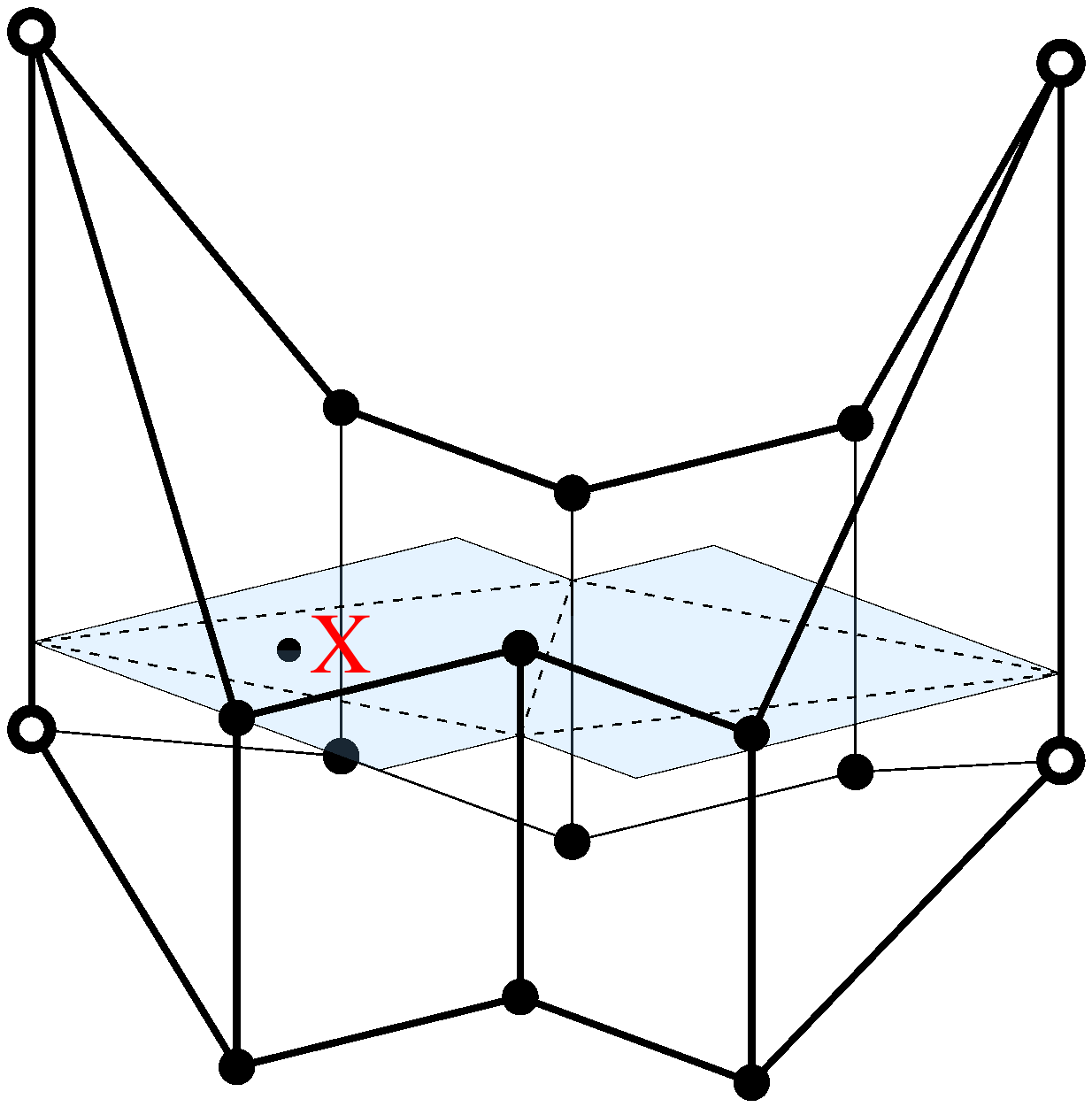} 
      }
    &
    \subfigure[][]{
      \label{fig:2edge:3D6C}
      \includegraphics[height=2.8cm]{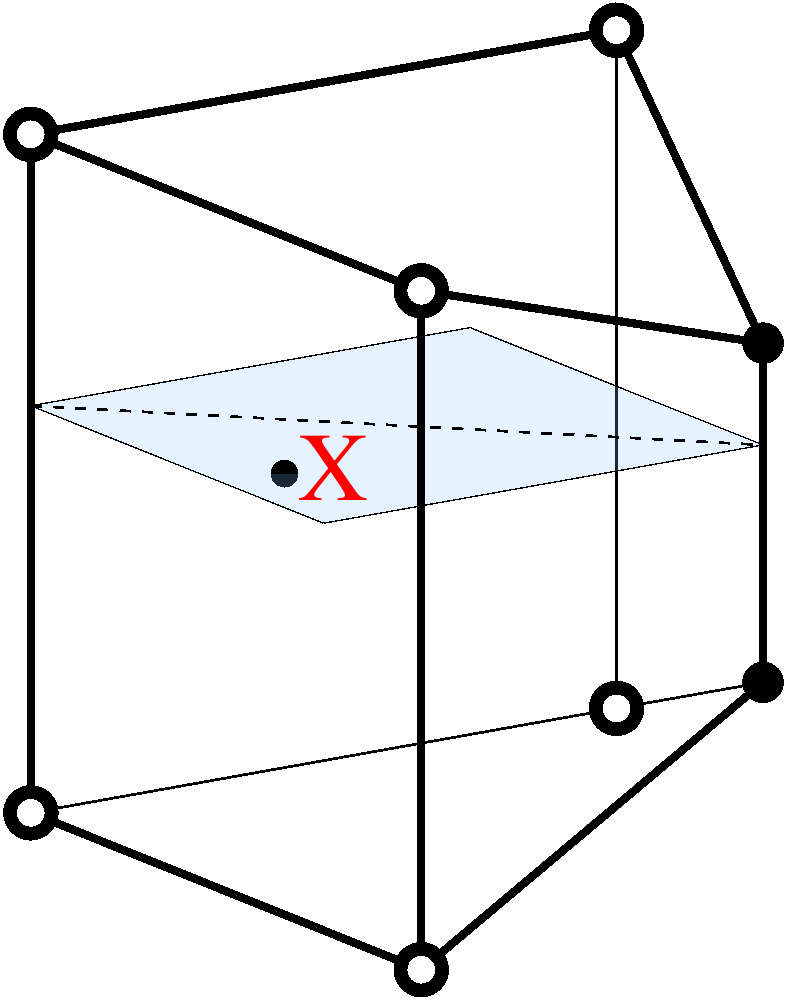} 
      }
    &
    \subfigure[][]{
      \label{fig:2edge:2D2C2}
      \includegraphics[height=3.5cm]{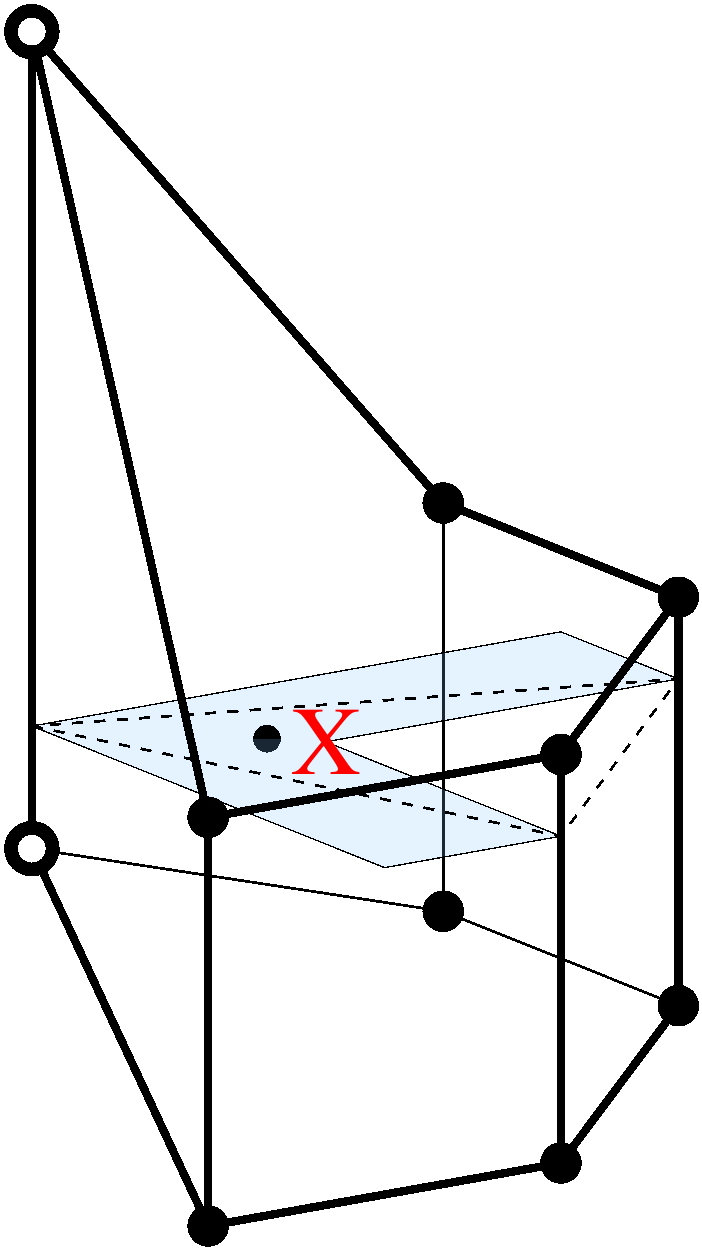} 
      }
    \end{tabular}
  }
    \caption{CSB decomposition (top panels only)
             and interpolation on 2-edge for 2D (top panels) 
             and 3D (bottom panels). 
             Herewith, hollow points are centers of Coarser cells, 
             solid points are those of Finer cells.
             Here, bolder solid points 
             (panels \subref{fig:2edge:2D1C}-\subref{fig:2edge:2D3C}) mark
             Finer vertices of a central quadrangle.
             Shaded regions show main interpolation planes 
             (which become resolution 2-edges in 2D) 
             for point {\bf X} (marked for 3D),
             white regions in top panels are resolution 0- and 1-edges.
             CSB decomposition is shown with red lines, 
             being dividing lines.
             Triangulation of the main interpolation plane 
             is shown with dashed lines.
             It is easy to see from panels 
             \subref{fig:2edge:2D1C}-\subref{fig:2edge:2D3C}
             that interpolation on resolution 2-edges
             continuously transits to interpolation 
             on resolution 0- and 1-edges
             through boundaries of main interpolation planes.
             Note, that certain triangles in the main interpolation plane
             only partially lay inside a resolution 2-edge
             with the remaining parts being inside resolution 1-edges.
            }
    \label{fig:2edge}
\end{figure}

%% file: CSB.3edge.tex
\input{fig.3edge}

If the edge type of the central hexahedron 
appears to be 3,
then, generally speaking, the CSB decomposes into
resolution 1-, 2- and 3-edges.
The approach for interpolation is similar to that applied
in the previous section.
Now, the dividing planes are introduced in three directions.
We sort out domains, which are resolution 1- and 2-edges.
Specifically, if one of the CSB faces is fully Coarse,
then the rectangular box confined between this face and a Fine 4-cluster
is a resolution 1-edge 
(see panel \subref{fig:sortoutedge:1} of Figure \ref{fig:sortoutedge}).
Similarly, in configuration presented 
in panel \subref{fig:sortoutedge:2} of Figure \ref{fig:sortoutedge},
with one of the edges of CSB being fully Coarse,
several rectangular boxes may be found, 
which form a domain of the resolution 2-edge with the trivial
elements parallel to an isolated Coarse 2-cluster.
This can be derived by analyzing the level set
and decomposition of the CBS into rectangular boxes.
If the point falls into a resolution 1-edge or 
a resolution 2-edge, the interpolation is performed
as described above for such resolution edges 
in section \ref{sec:1edge} and \ref{sec:2edge}, respectively 
and the algorithm quits.

If the point {\bf X} falls into the central box, 
i.e. the enclosed set of a central shape,
then interpolation is performed 
using the tessellation shown in Figure \ref{fig:3edge}.
In this case the interpolation procedure is the same 
as that in \cite{Weber2001DataVis}.
The algorithm quits.

\input{fig.sortoutedge}

Finally, for the point {\bf X}, which doesn't fall into either
the set enclosed by central hexahedron, or resolution 1- and 2-edges,
the interpolation stencil is chosen according to 
resolution 3-edge decomposition outside the central hexahedron
into simpler shapes. 
The interpolation is performed 
on the resulting shapes.
The particular pattern of decomposition is based
on the current CSB configuration and the position of 
the rectangular box {\bf X} falls into.
The idea of the decomposition procedure 
is given in Figure \ref{fig:edgeboundary}.

\input{fig.edgeboundary}

At this point another branch in the CSB decomposition procedure is possible,
as long as point {\bf X} may happen to actually fall 
into the central hexahedron,
rather than into the constructed shape,
thus compromising the effort spent for its construction and decomposition.
In this case the interpolation is performed on the central hexahedron
as described in Figure \ref{fig:3edge}.
The advantage of our resolution edge concept is that
it allows us to avoid using whenever possible this 
complicated branch in the algorithm for points
falling into resolution 1- and 2-edges.

We emphasize, that at this point our algorithm 
becomes more complicated than that in \cite{Weber2001DataVis},
though interpolation procedures for points 
within resolution 3-edge are the same.
However, we benefit from the interpolation procedure being simpler
on a much larger part of the grid (resolution 1- and 2-edges),
which dominate the computational time.
Therefore, complication of the algorithm itself and its implementation
is merely a trade-off for a higher efficiency.

%% file: fig.3edge.tex
\begin{figure}
  \renewcommand{\thesubfigure}{\Alph{subfigure}}
  \makebox[\linewidth][c]{
    \begin{tabular}{p{3.1cm} p{3.1cm} p{3.1cm} p{3.1cm}}
      \subfigure[][]{
        \label{fig:3edge:11}
        \includegraphics[height=2.25cm]{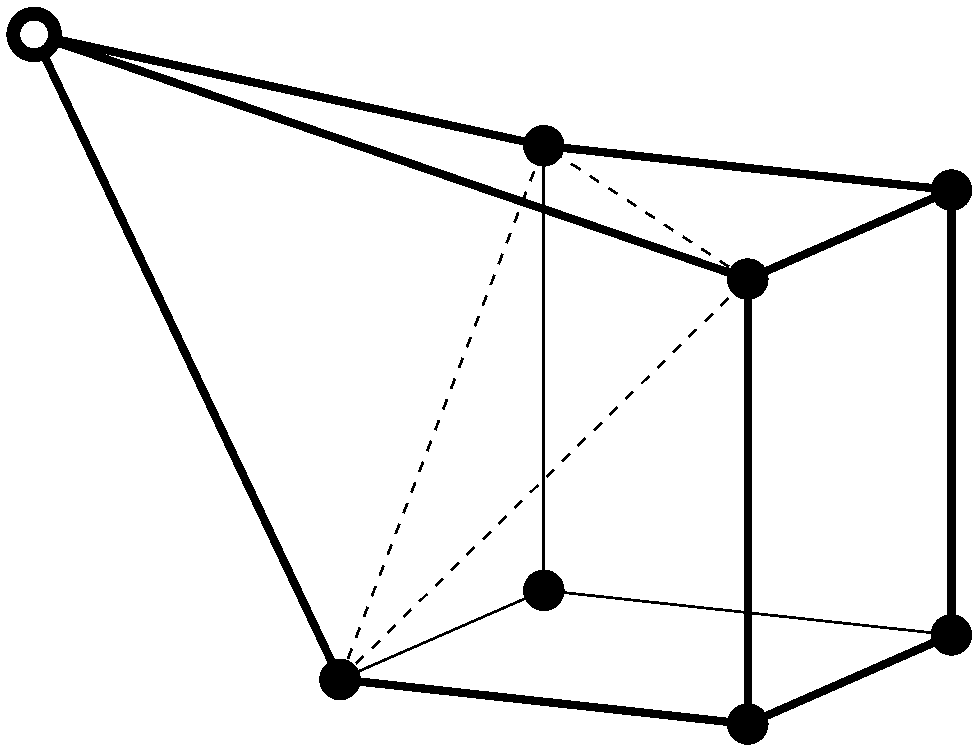}     
        }
      &
      \subfigure[][]{
        \label{fig:3edge:12}
        \includegraphics[height=3.0cm]{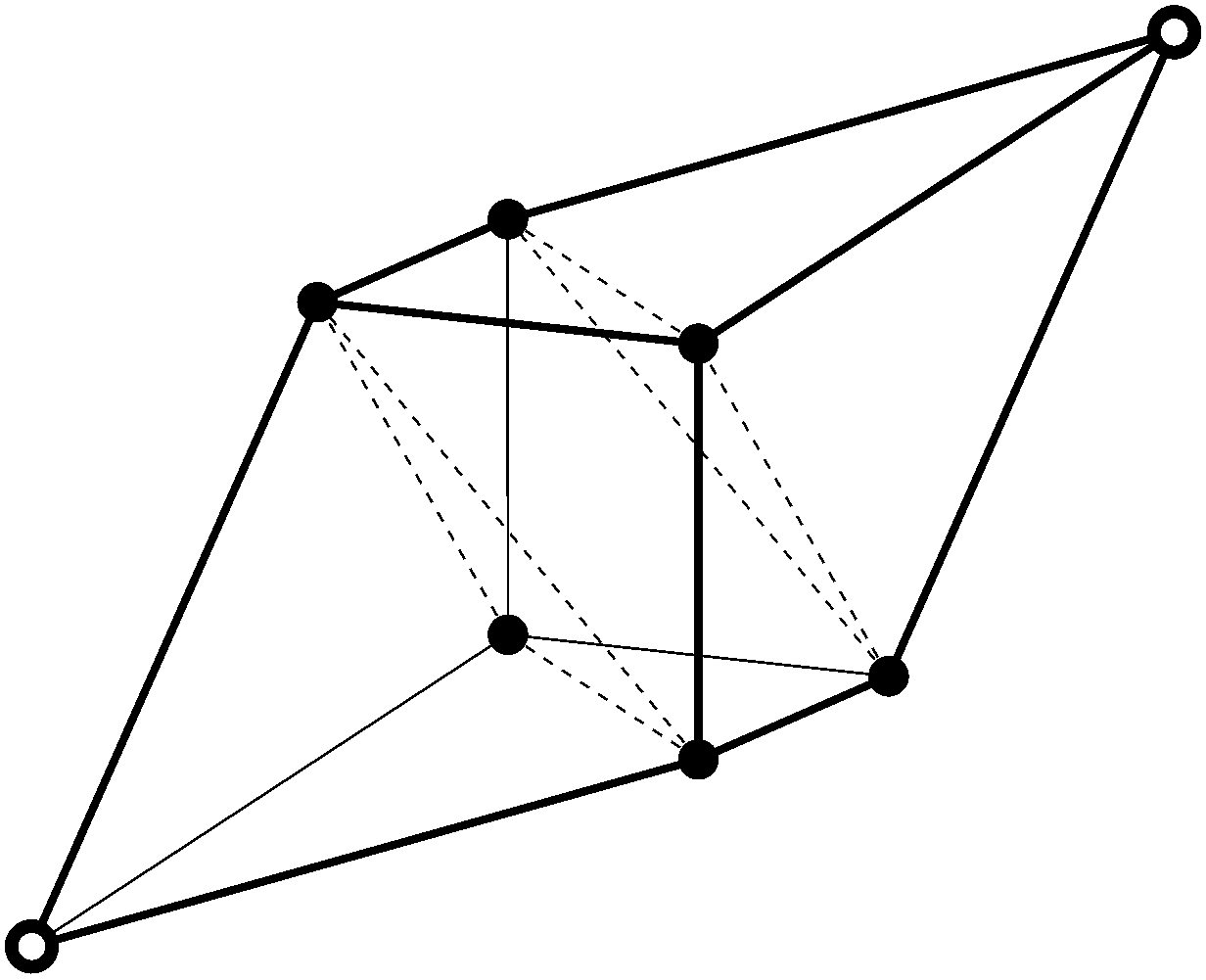}      
        }
      &
      \subfigure[][]{
        \label{fig:3edge:13}
        \includegraphics[height=2.25cm]{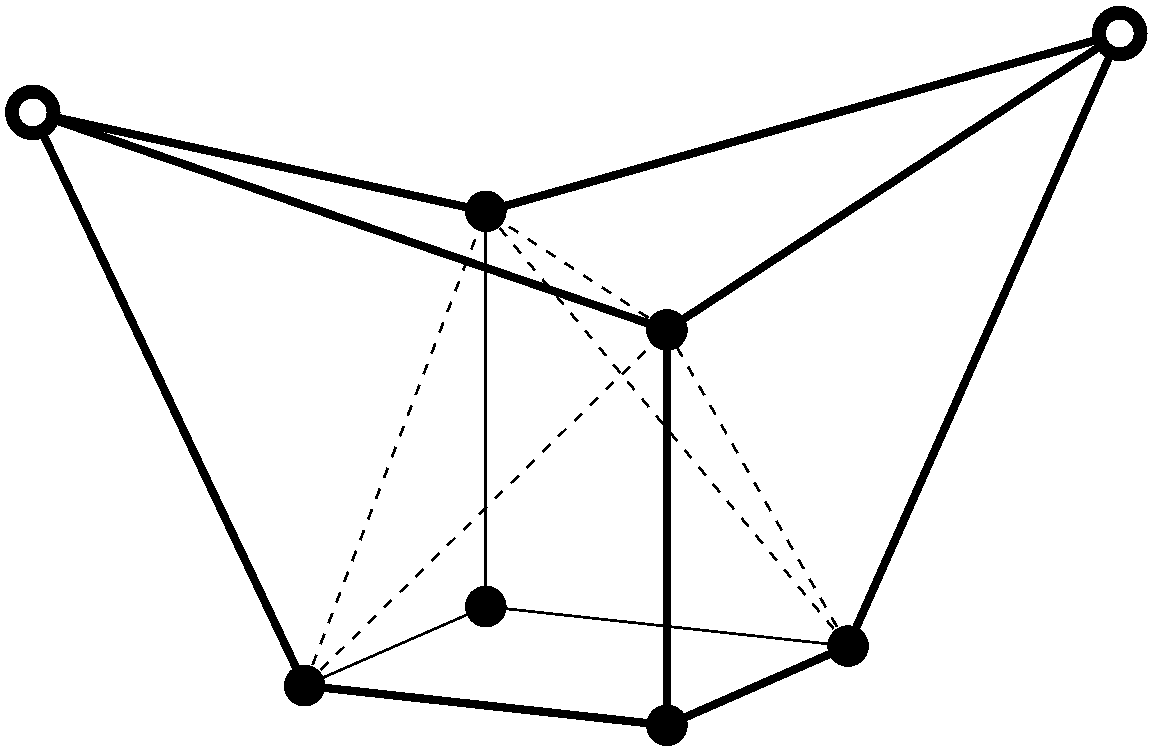} 
        }
      &
      \subfigure[][]{
        \label{fig:3edge:14}
        \includegraphics[height=2.25cm]{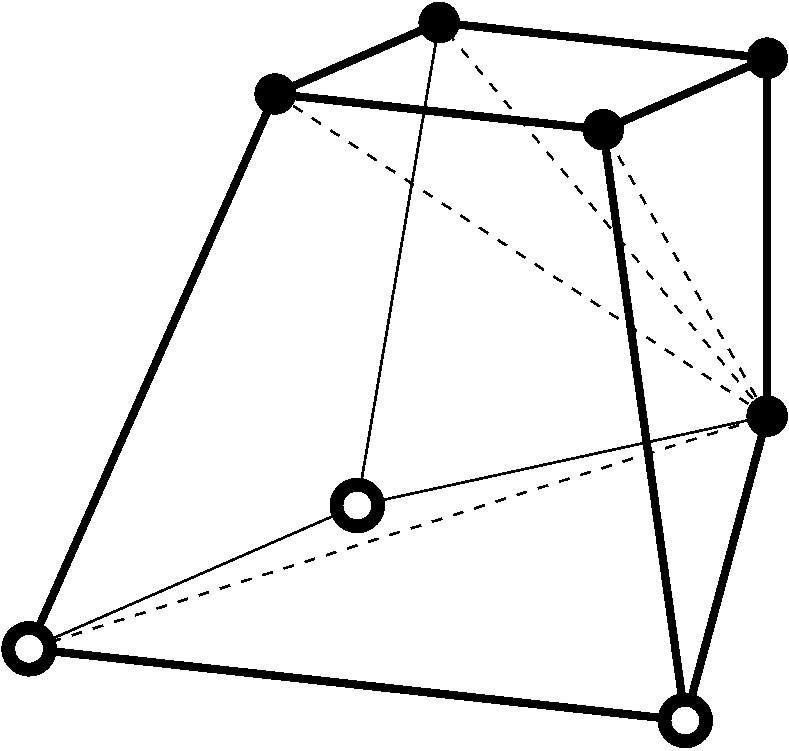}       
        }
      \\[-9pt]
      \subfigure[][]{
        \label{fig:3edge:21}
        \includegraphics[height=3.0cm]{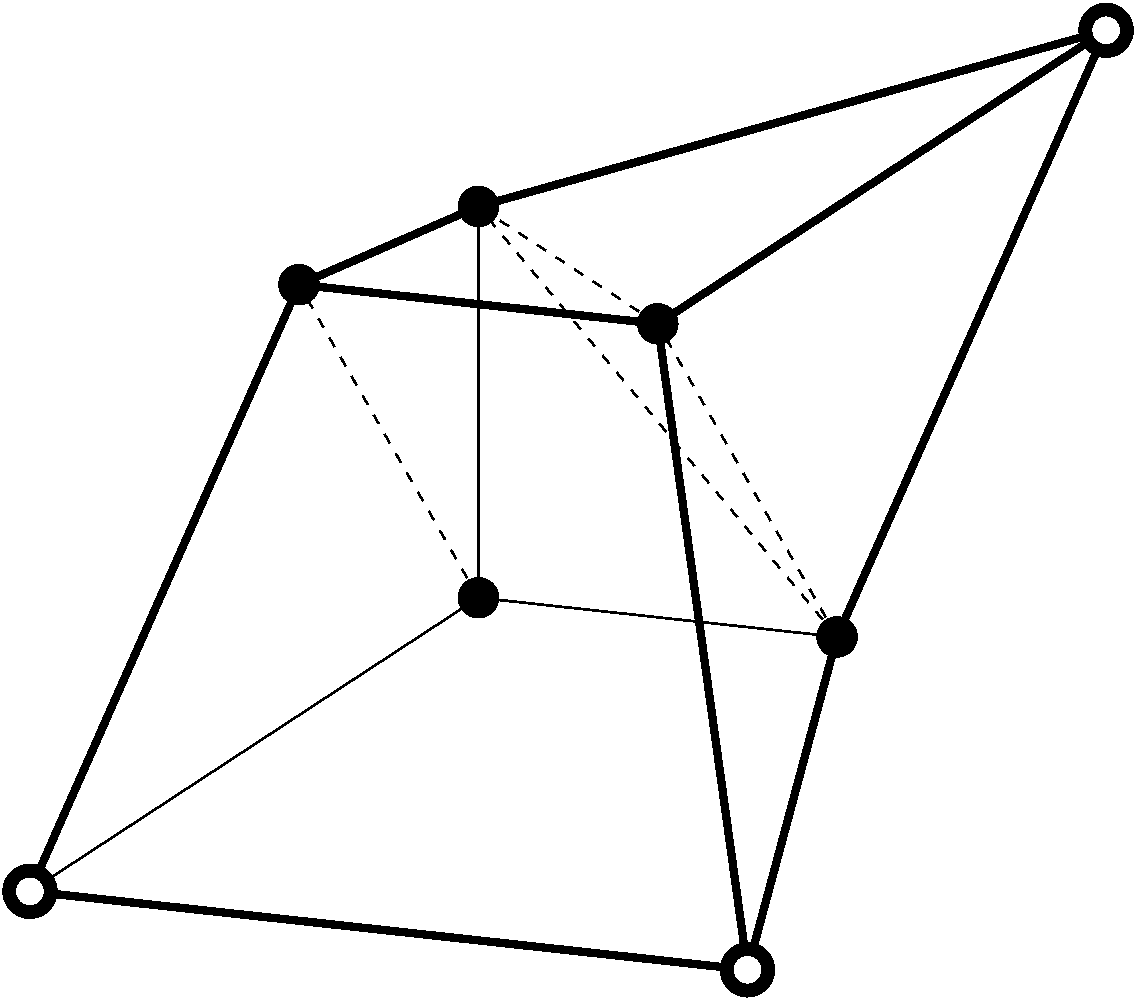}  
        }
      &
      \subfigure[][]{
        \label{fig:3edge:22}
        \includegraphics[height=3.0cm]{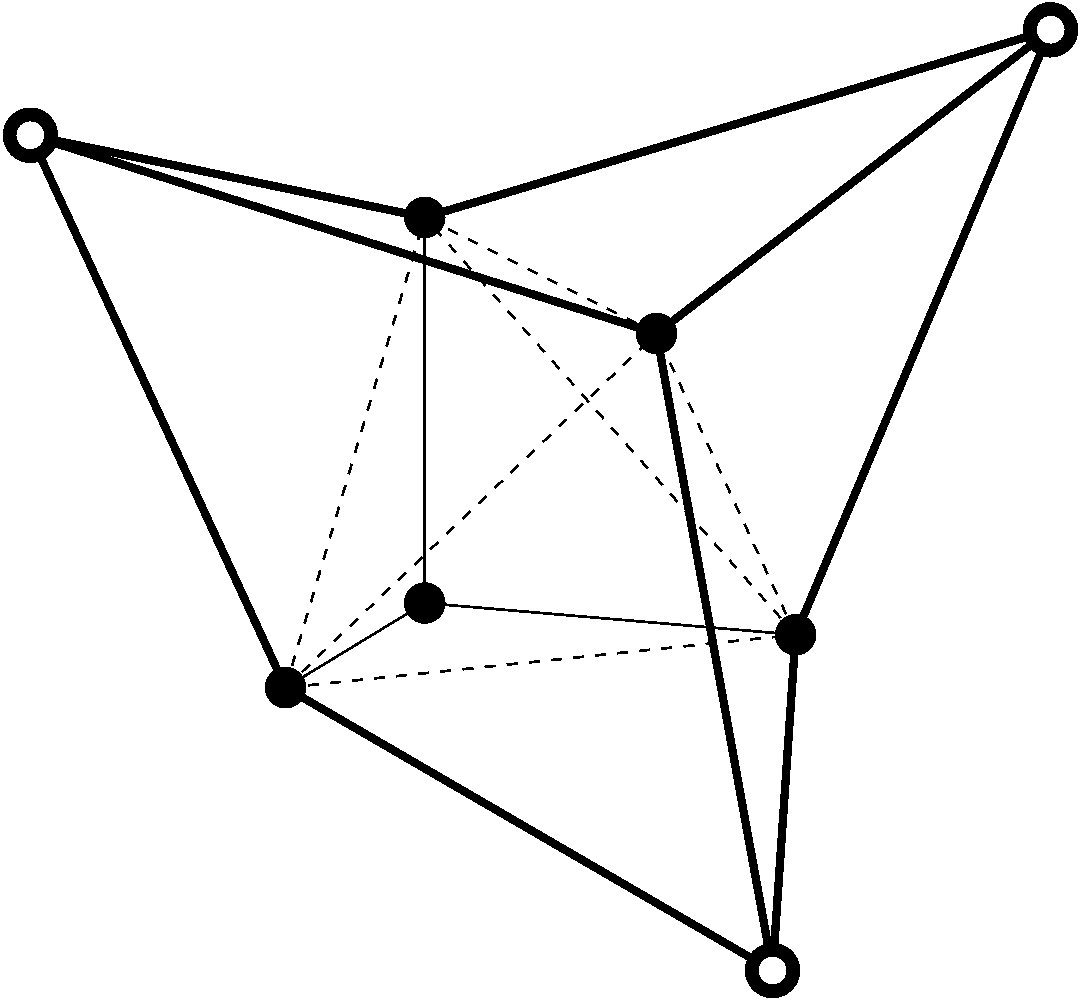}  
        }
      &
      \subfigure[][]{
        \label{fig:3edge:23}
        \includegraphics[height=3.0cm]{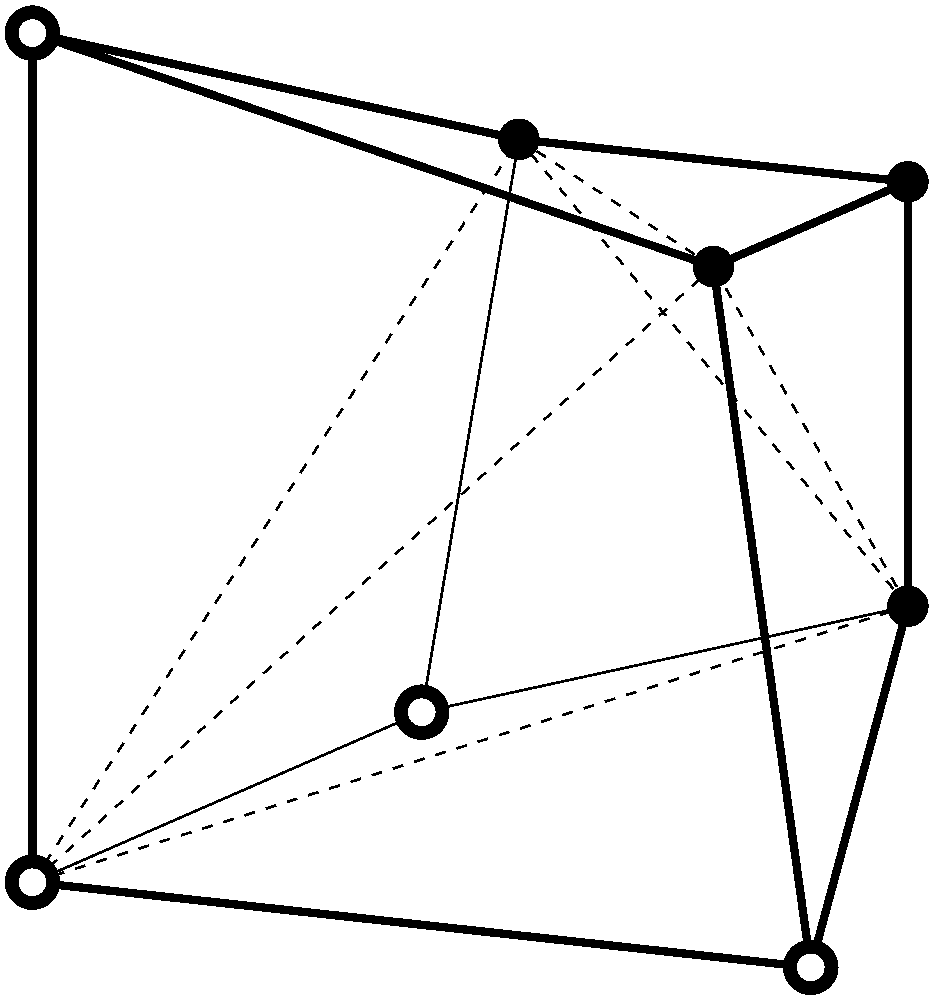}      
        }
      &
      \subfigure[][]{
        \label{fig:3edge:24}
        \includegraphics[height=3.0cm]{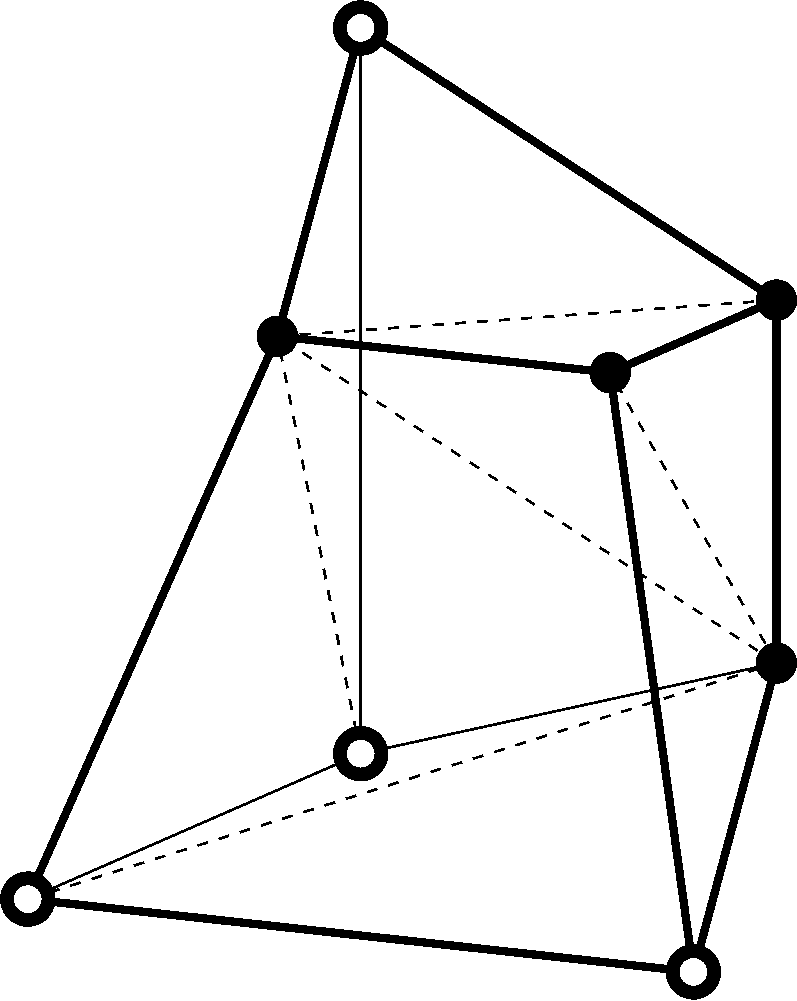}  
        }
      \\[-9pt]      
      \subfigure[][]{
        \label{fig:3edge:31}
        \includegraphics[height=3.0cm]{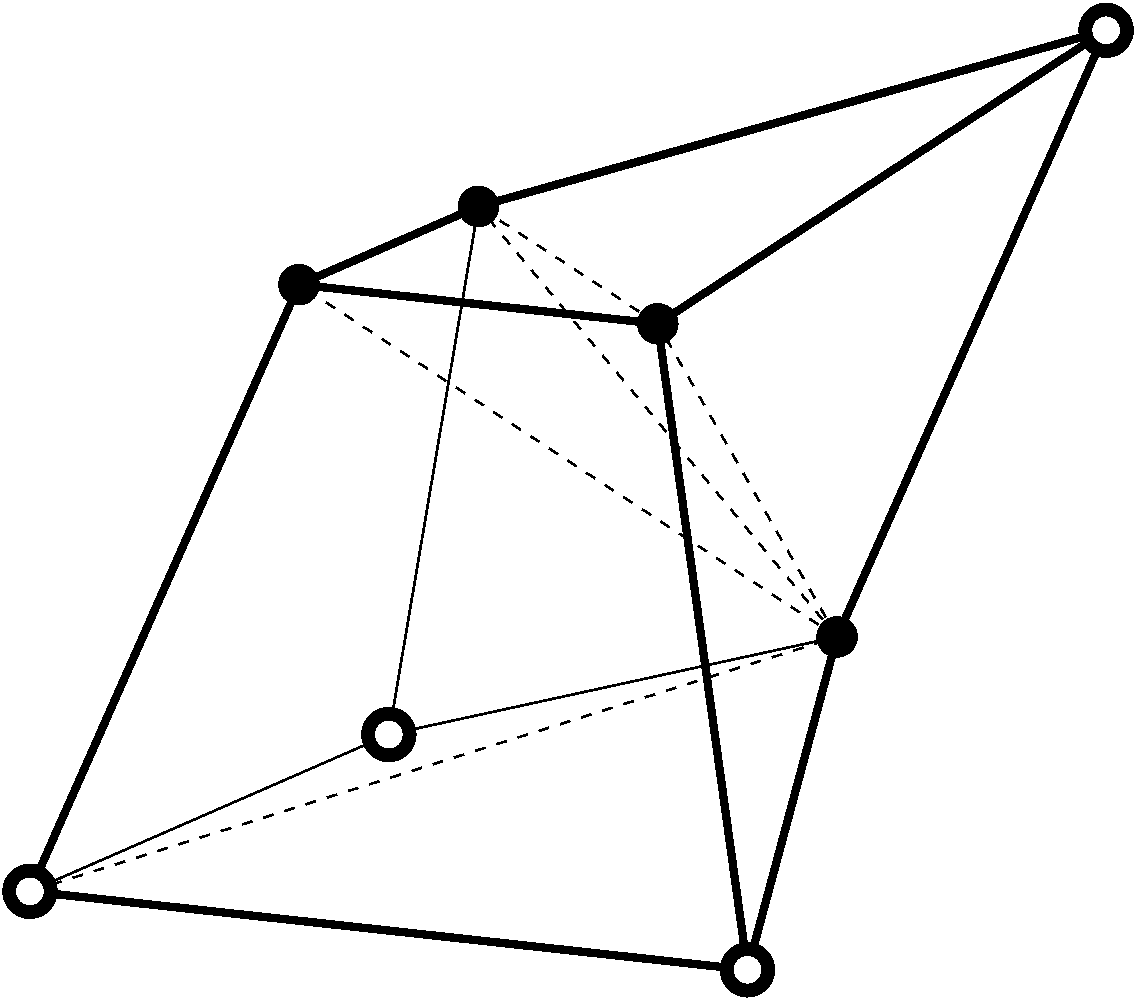} 
        }
      &
      \subfigure[][]{
        \label{fig:3edge:32}      
        \includegraphics[height=3.0cm]{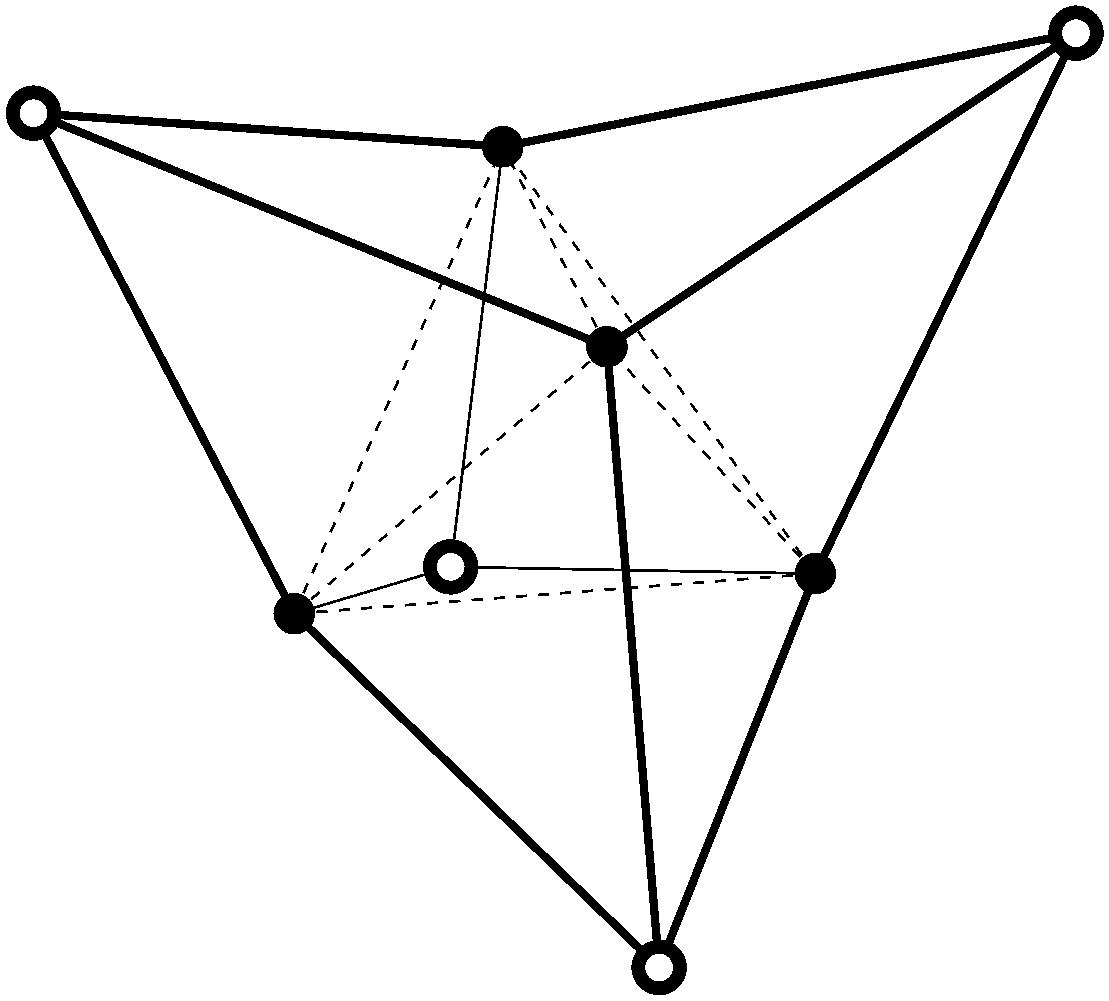}  
        }
      &
      \subfigure[][]{
        \label{fig:3edge:33}
        \includegraphics[height=3.0cm]{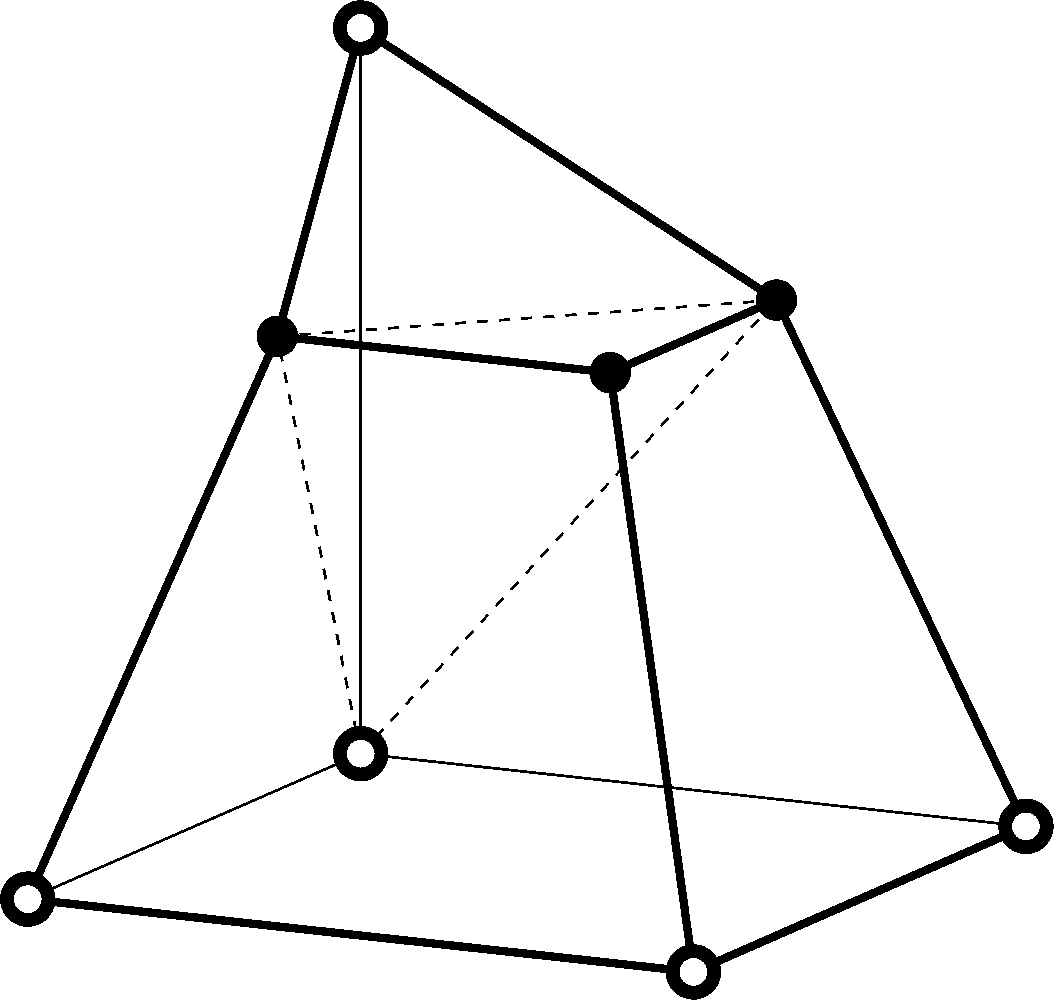}      
        }
      &
      \subfigure[][]{
        \label{fig:3edge:34}
        \includegraphics[height=3.0cm]{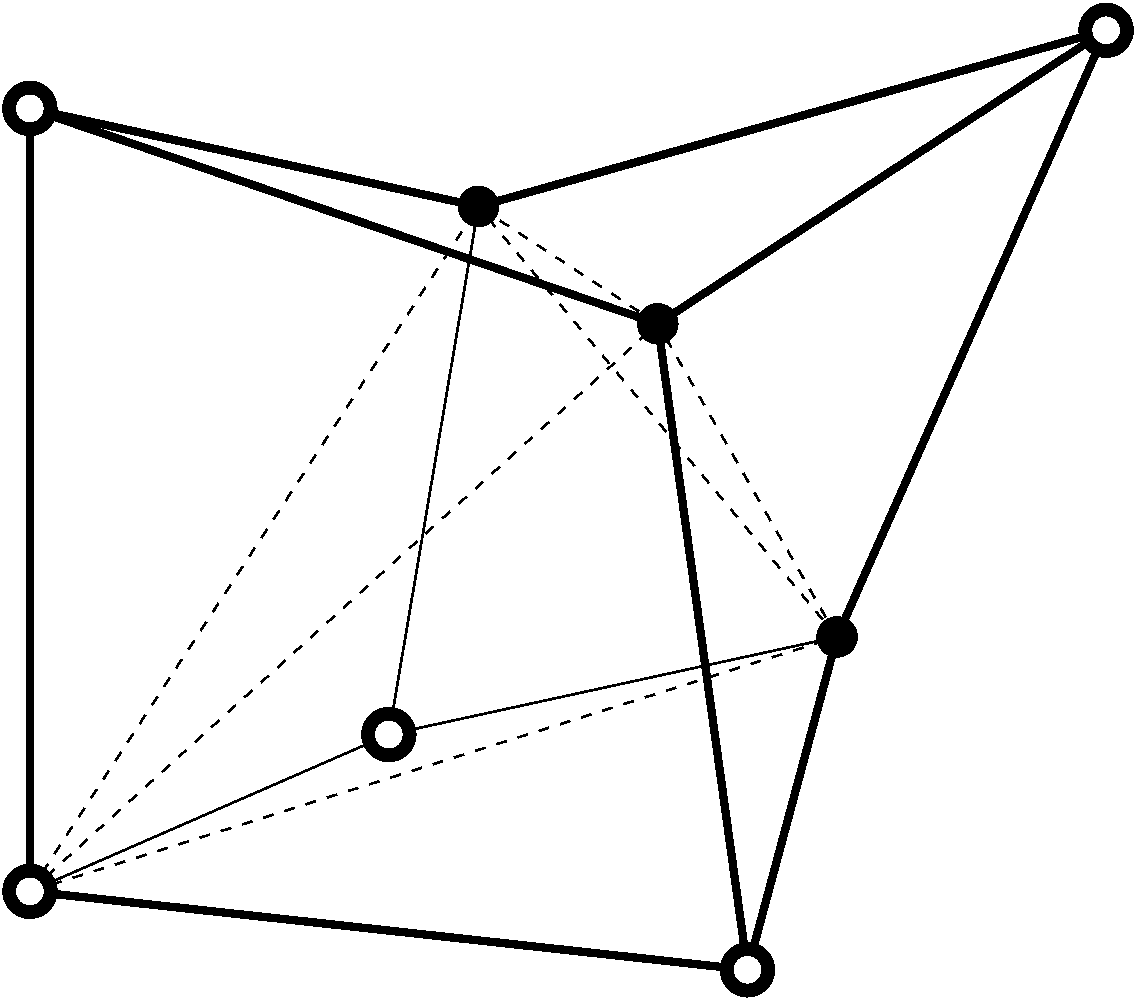}   
        }
      \\[-9pt]      
      \subfigure[][]{
        \label{fig:3edge:41}      
        \includegraphics[height=3.0cm]{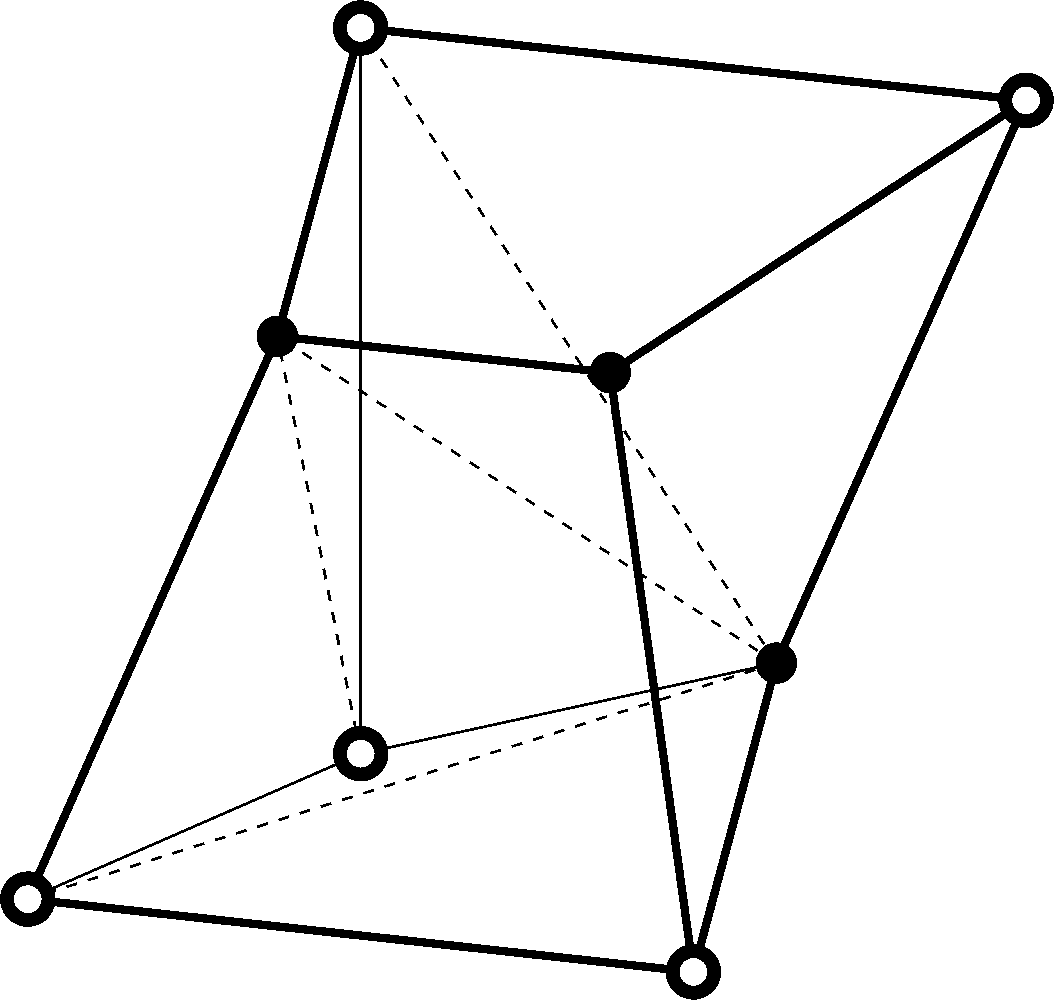} 
        }
      &
      \subfigure[][]{
        \label{fig:3edge:42}      
        \includegraphics[height=3.0cm]{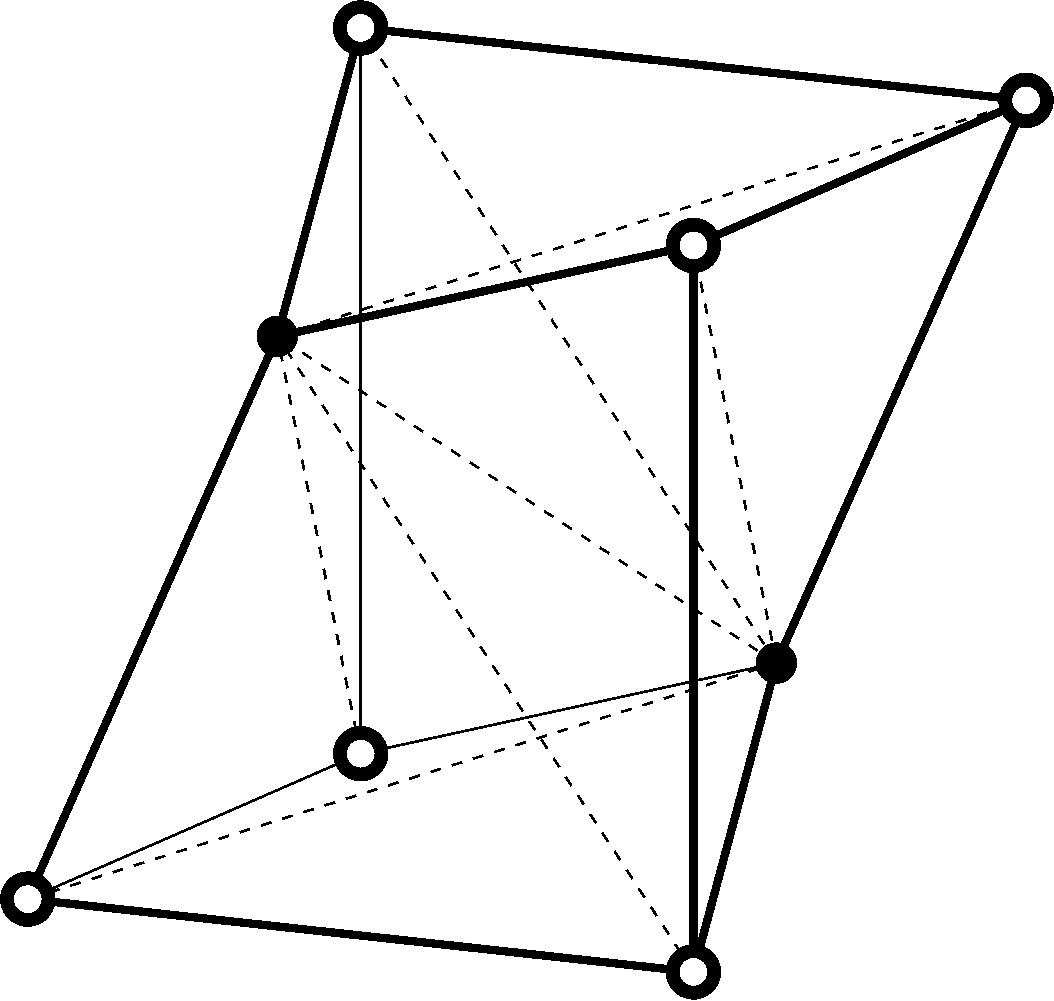}      
        }
      &
      \subfigure[][]{
        \label{fig:3edge:43}
         \includegraphics[height=3.0cm]{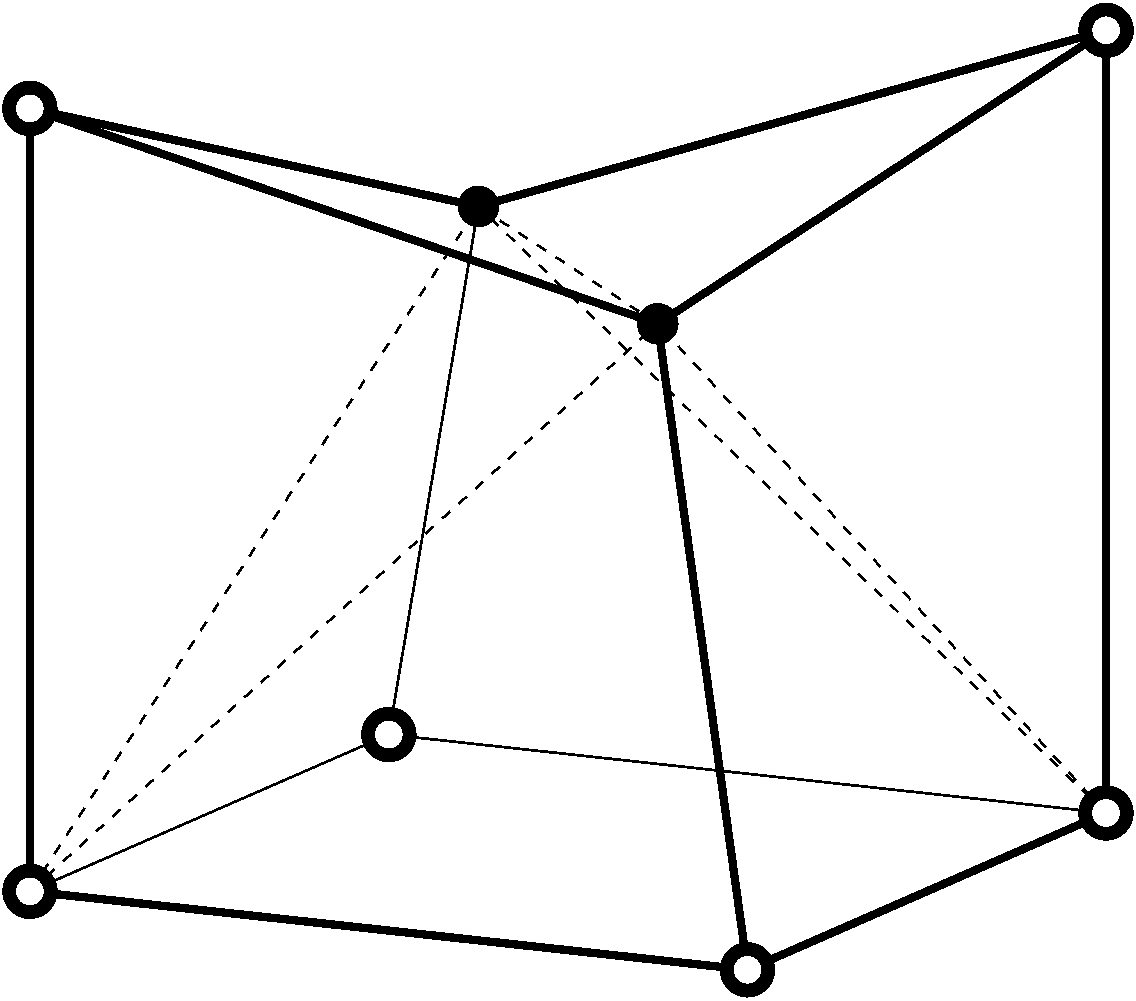}  
         }
      &
      \subfigure[][]{
        \label{fig:3edge:44}
        \includegraphics[height=3.0cm]{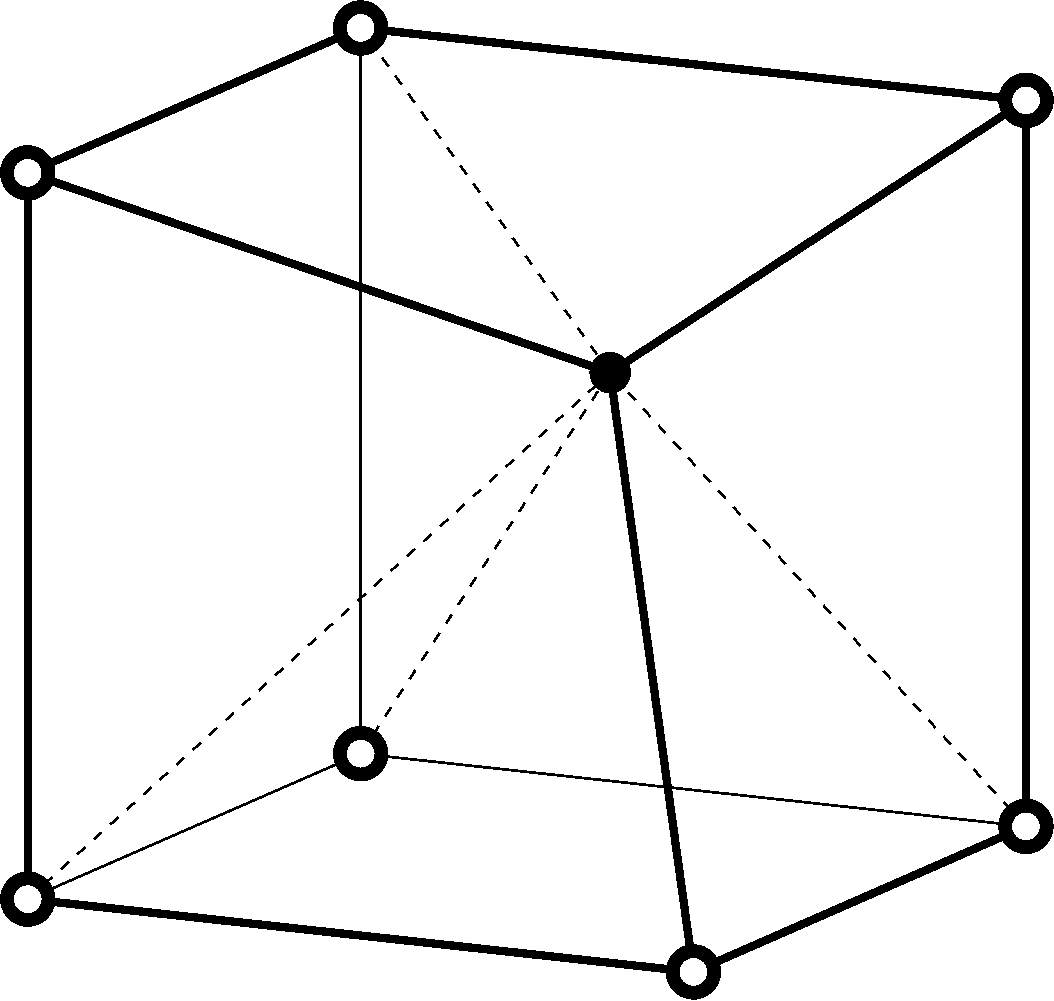}
        }
    \end{tabular}
  }
    \caption{Decomposition of central hexahedra of resolution 3-edges.
             Particular decomposition for each hexahedron is shown with dashed lines:
              \subref{fig:3edge:11} - 
                  tetrahedron  and irregular shape;
              \subref{fig:3edge:12} -  
                  2 tetrahedra and irregular shape;
              \subref{fig:3edge:13} - 
                  2 tetrahedra and irregular shape;
              \subref{fig:3edge:14} - 
                  rectangular and 2 trapezoidal pyramids;
              \subref{fig:3edge:21} - 
                  tetrahedron, rectangular pyramid and wedge;
              \subref{fig:3edge:22} - 
                  5 tetrahedra;
              \subref{fig:3edge:23} - 
                  5 tetrahedra;
              \subref{fig:3edge:24} - 
                  2 tetrahedra and 2 trapezoidal pyramids;
              \subref{fig:3edge:31} - 
                  2 tetrahedra and 2 trapezoidal pyramids;
              \subref{fig:3edge:32} - 
                  5 tetrahedra;
              \subref{fig:3edge:33} - 
                  tetrahedron and irregular shape; 
              \subref{fig:3edge:34} - 
                  5 tetrahedra;
              \subref{fig:3edge:41} - 
                  2 tetrahedra and 2 trapezoidal pyramids;
              \subref{fig:3edge:42} - 
                  6 tetrahedra;
              \subref{fig:3edge:43} - 
                  2 tetrahedra and irregular shape;
              \subref{fig:3edge:44} - 
                  3 rectangular pyramids.}
    \label{fig:3edge}
\end{figure}

%% file: fig.sortoutedge.tex
\begin{figure}[!h]
  \centering
  \renewcommand{\thesubfigure}{\Alph{subfigure}}
  \begin{tabular}{cc}
    \subfigure[][]{
      \label{fig:sortoutedge:1}
      \includegraphics[height=4.0cm]{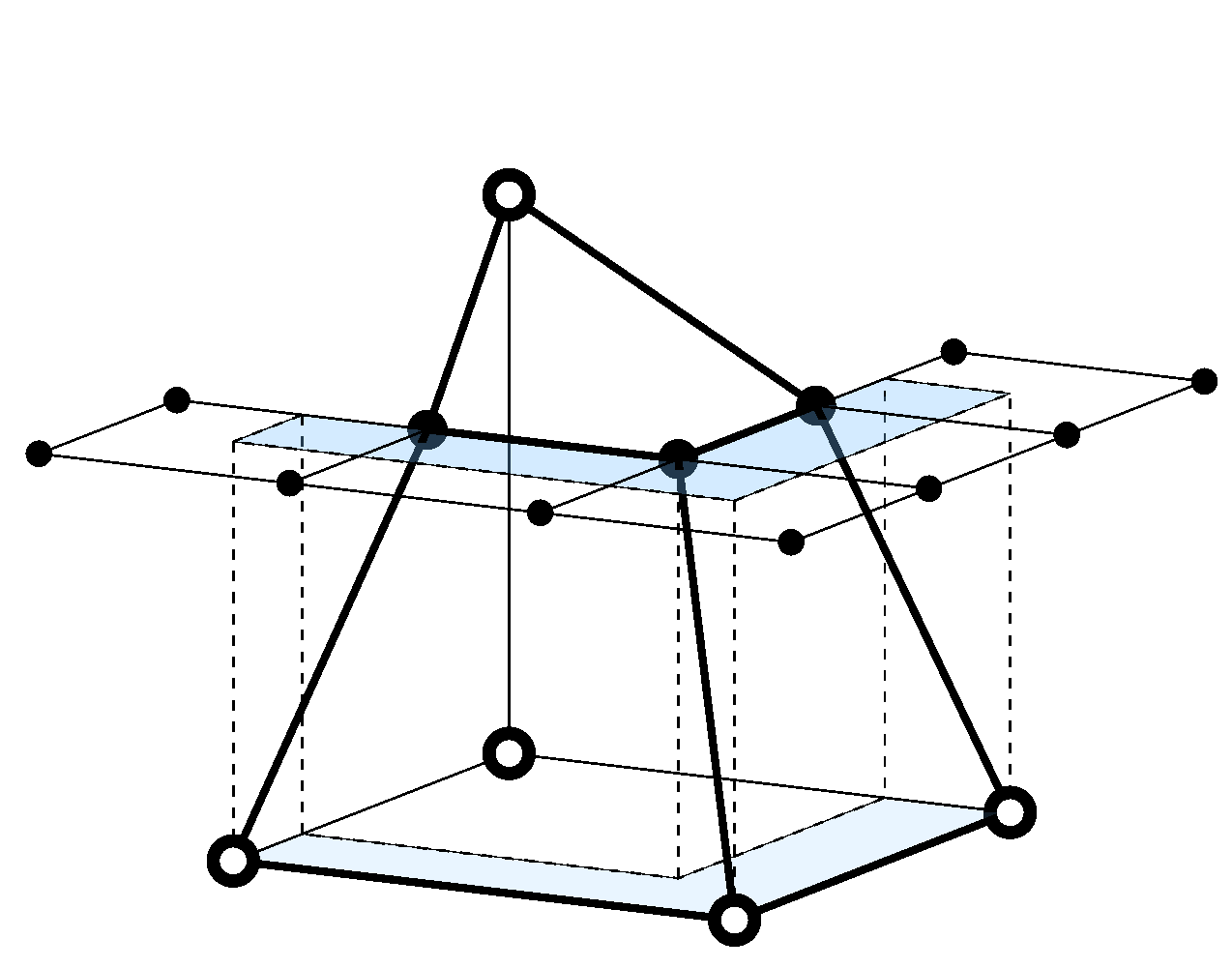}
      } &
    \subfigure[][]{
      \label{fig:sortoutedge:2}
      \includegraphics[height=4.0cm]{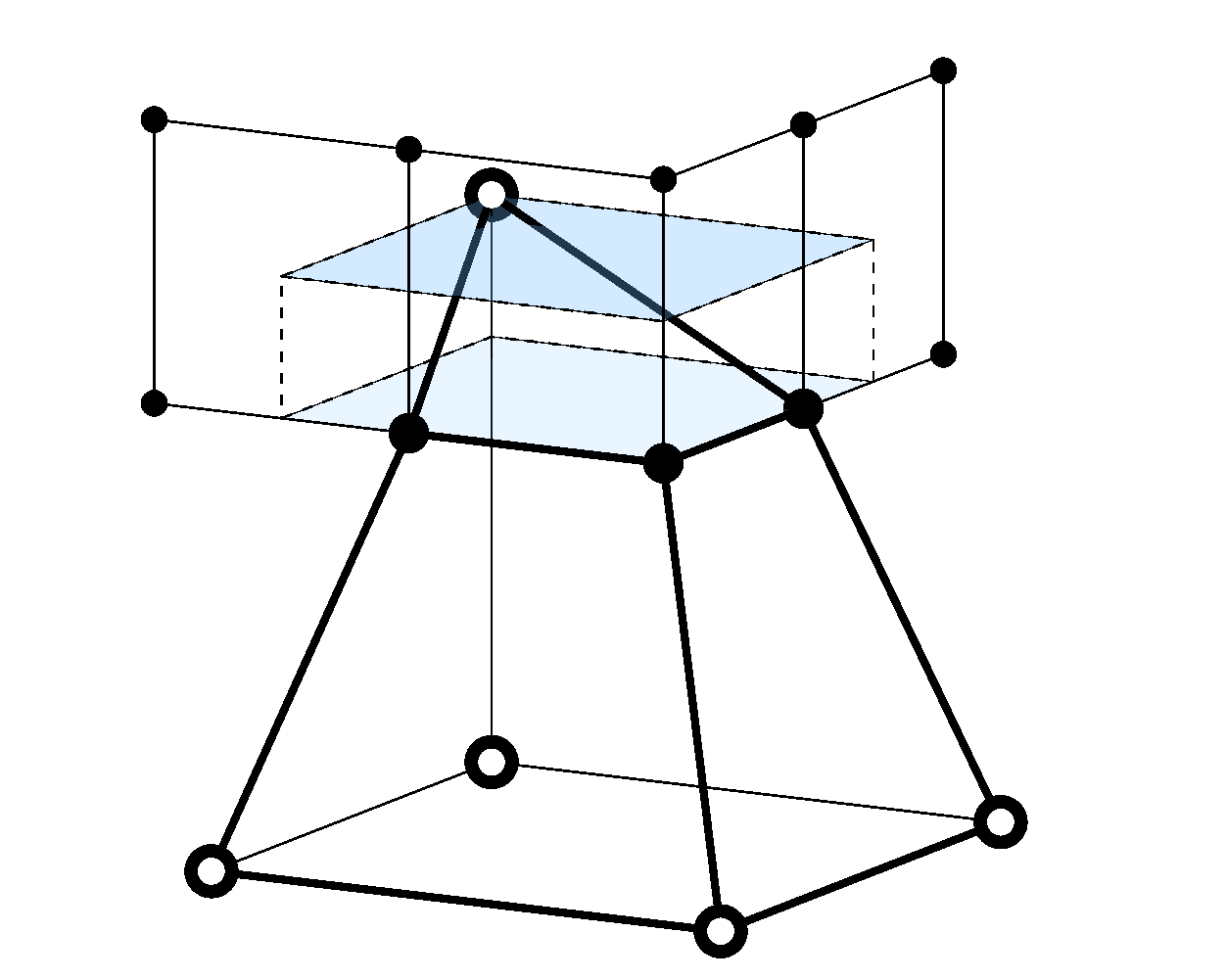}
      } 
  \end{tabular}
  \caption{An example of a CSB including resolution 1- and 2-edges. 
           Within the CSB with central octagon \subref{fig:3edge:33} in Figure \ref{fig:3edge}
           there are domains, which are: 
           \subref{fig:sortoutedge:1} - a resolution 1-edge;
           \subref{fig:sortoutedge:2} - a resolution 2-edge.
           Borders of these domains are shown with dashed lines,
           their top and bottom faces are shaded.
  }
  \label{fig:sortoutedge}
\end{figure}

%% file: fig.edgeboundary.tex
\begin{figure}[!h]
  \renewcommand{\thesubfigure}{\Alph{subfigure}}
  \makebox[\linewidth][c]{
    \begin{tabular}{cc}
      \subfigure[][]{
        \label{fig:edgeboundary:1C}
        \includegraphics[height=3.5cm]{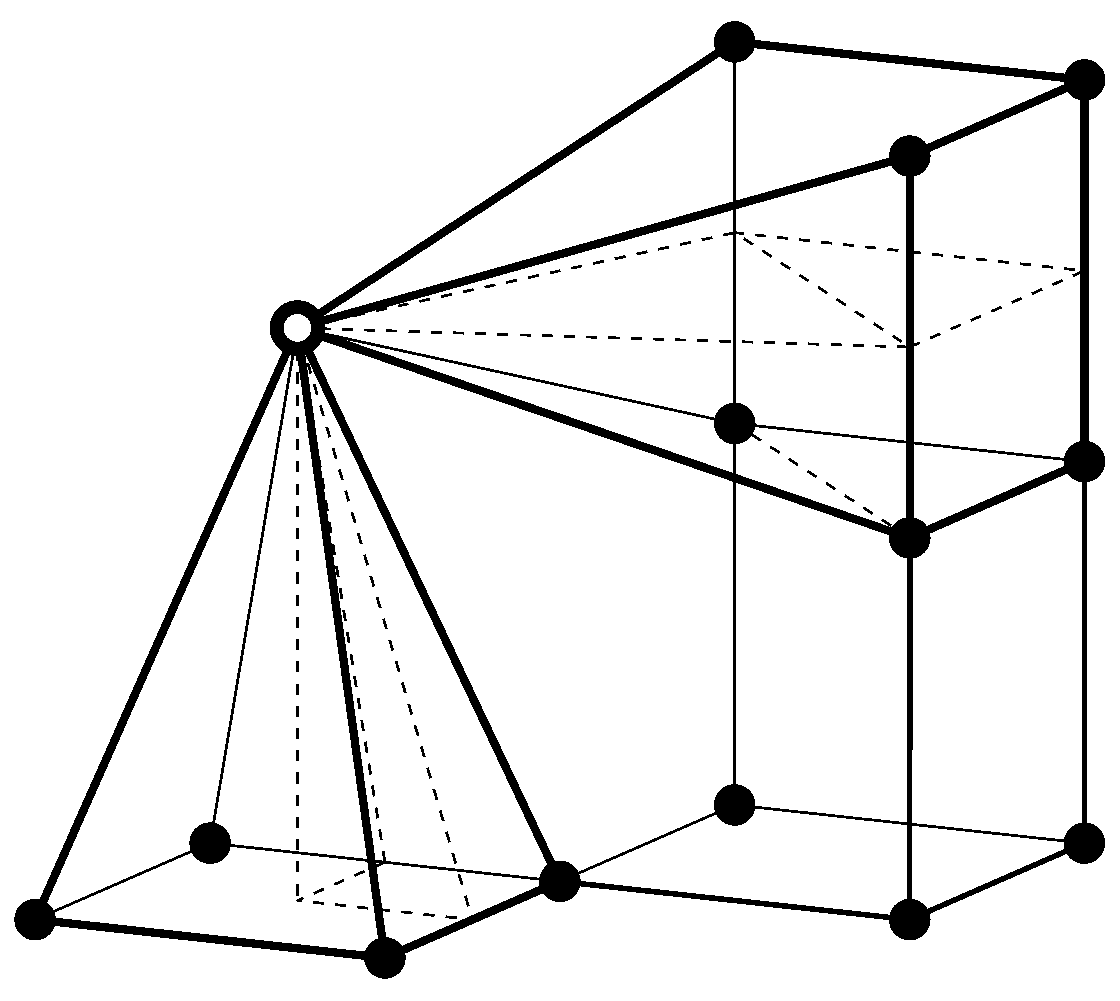}
        }
      &
      \subfigure[][]{
        \label{fig:edgeboundary:2C2}
        \includegraphics[height=3.5cm]{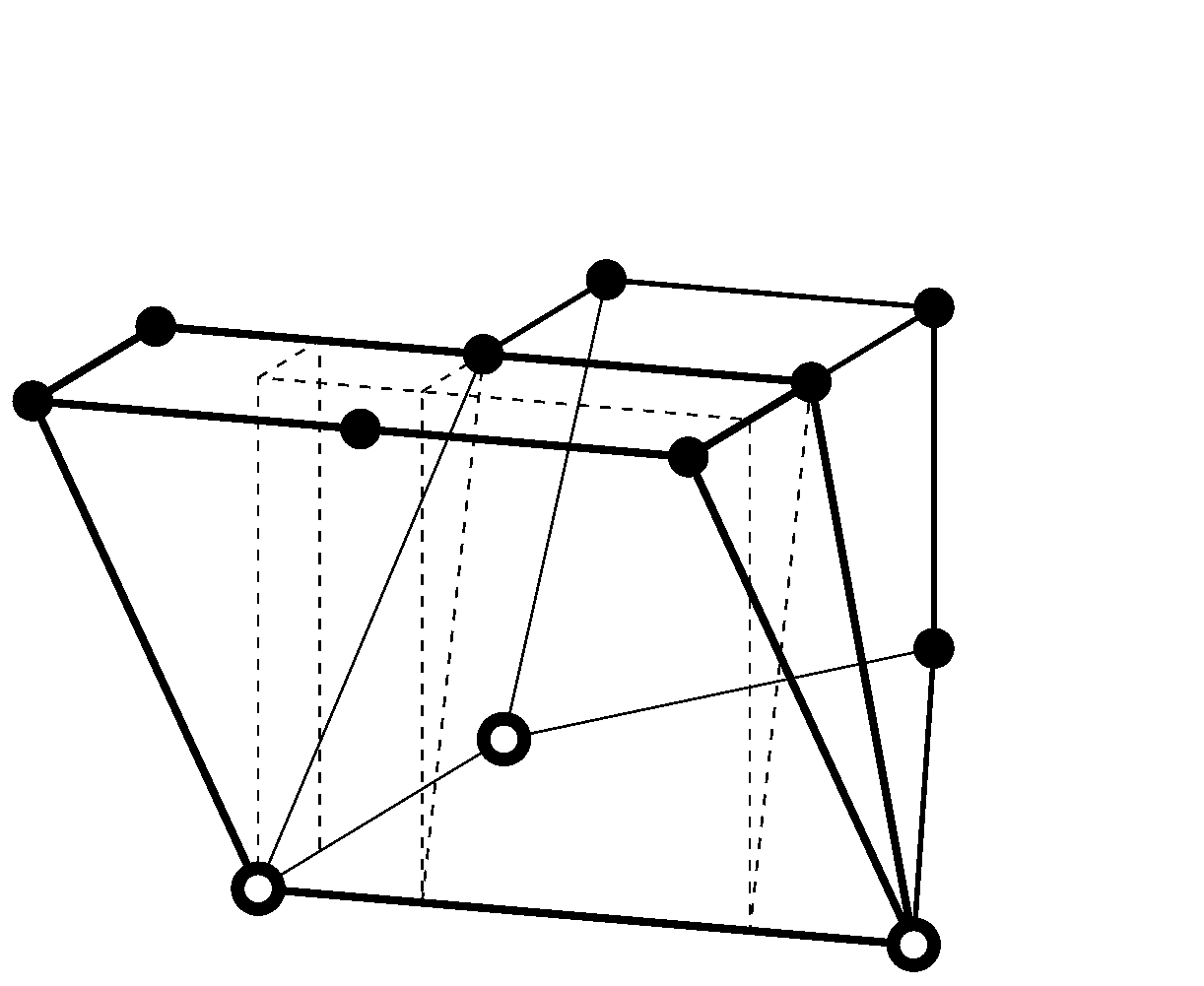}
        }
      \\[-8pt]
      \subfigure[][]{
        \label{fig:edgeboundary:2C1}
        \includegraphics[height=3.5cm]{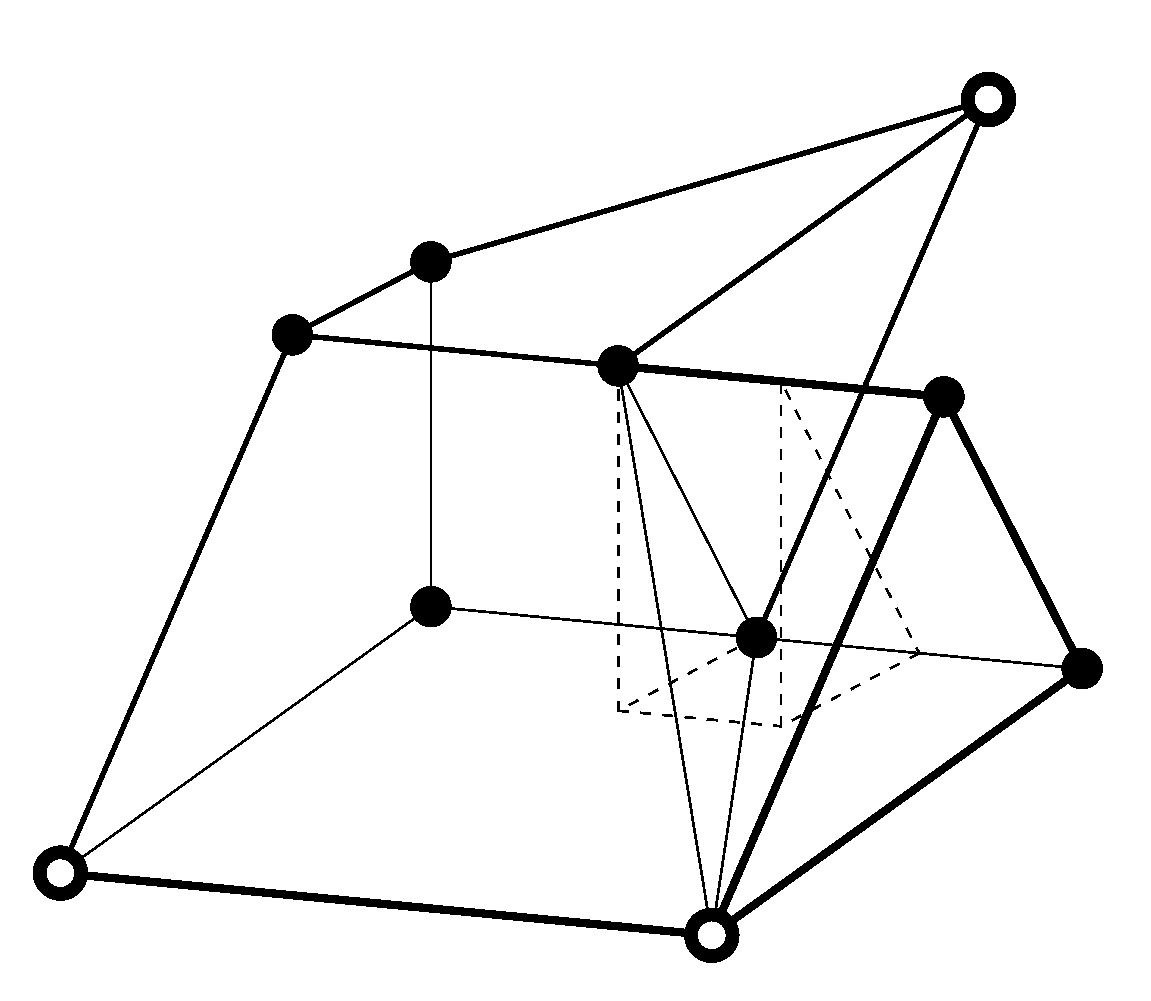}
        }
      &
      \subfigure[][]{
        \label{fig:edgeboundary:4C}
        \includegraphics[height=3.5cm]{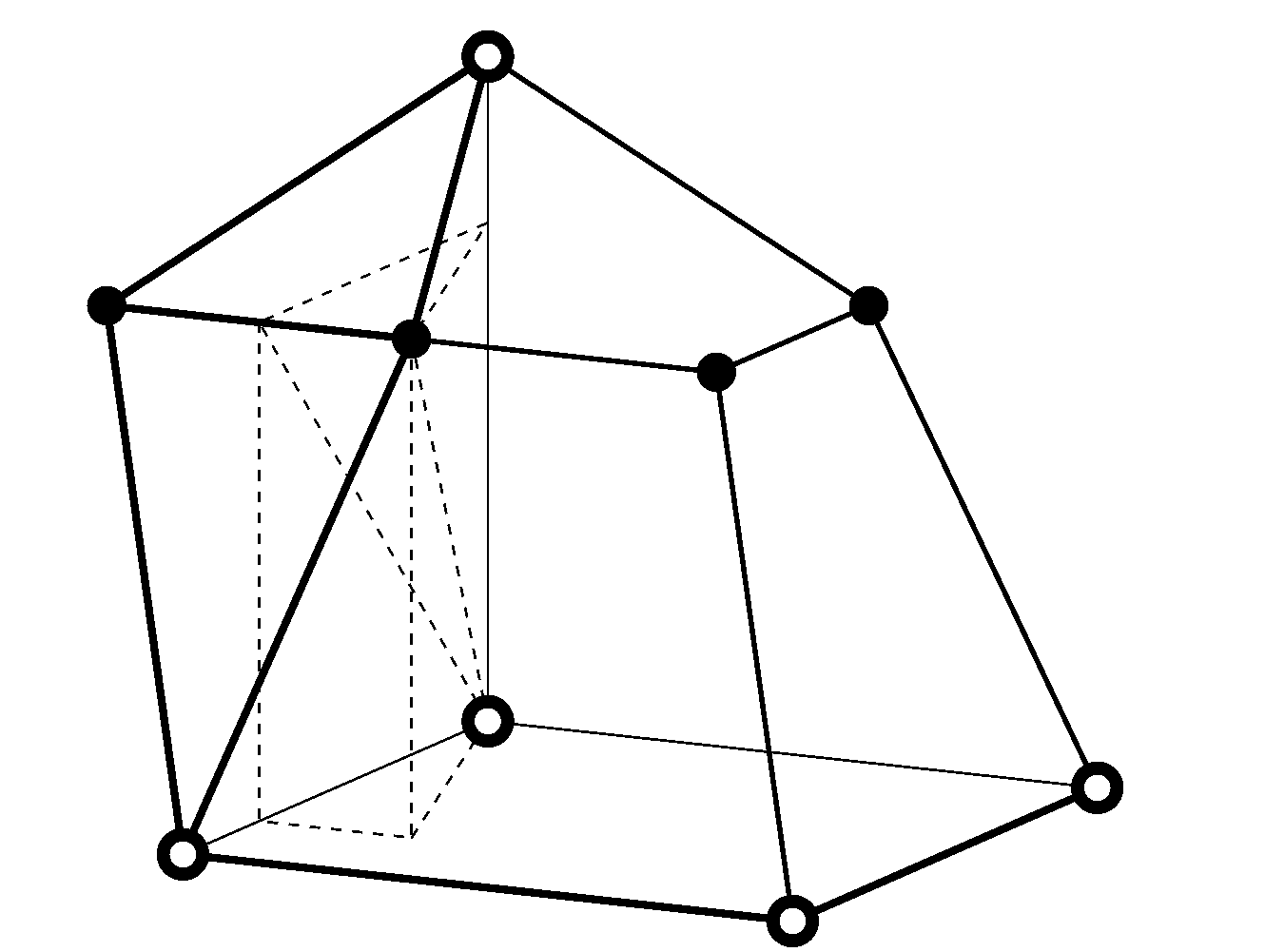}
        }
    \end{tabular}
    }
  \caption{Decomposition of a resolution 3-edge outside of 
           a central hexahedron into simple shapes.
           Dashed lines show the boundaries of these shapes.
           Panel \subref{fig:edgeboundary:1C} shows an isolated Coarse vertex,
           see panels \subref{fig:3edge:11}, 
           \subref{fig:3edge:12}, \subref{fig:3edge:13},
           \subref{fig:3edge:21}, \subref{fig:3edge:22}, 
           \subref{fig:3edge:31}, \subref{fig:3edge:32}
           and \subref{fig:3edge:34} in Figure \ref{fig:3edge},
           the following shapes may be distinguished: 
           a rectangular pyramid inside the one formed with 
           this Coarse vertex as an apex
           and either a Fine 4-cluster, 
           or a pair of Fine 2-clusters. 
           The latter may have a triangular prism formed 
           by 6 Fine vertices adjacent to it.
           Near fully Coarse edges,
           see panels \subref{fig:3edge:14}, \subref{fig:3edge:21}, 
           \subref{fig:3edge:23}, \subref{fig:3edge:24}, \subref{fig:3edge:31},
           \subref{fig:3edge:33}, 
           \subref{fig:3edge:34}, \subref{fig:3edge:41}, \subref{fig:3edge:42}
           and \subref{fig:3edge:43} in Figure \ref{fig:3edge}, 
           the following shapes may be  distinguished:
           (1) a triangular prism inside a triangular wedge formed 
           by a Coarse edge
           and either a Fine 4-cluster, or a pair of Fine 2-clusters 
           (panels \subref{fig:edgeboundary:2C2} 
           and \subref{fig:edgeboundary:2C1} respectively);
           (2) a tetrahedron inside the one formed by a Coarse edge  
           and a Fine 2-cluster
           (panel \subref{fig:edgeboundary:4C});
           (3) an irregular shape inside the one formed 
           by a pair of Coarse edges and a Fine 2-cluster
           (panel \subref{fig:edgeboundary:2C2}).
           Panel \subref{fig:edgeboundary:4C} shows 
           a fully Coarse face,
           see panels \subref{fig:3edge:33} 
           and \subref{fig:3edge:43}  in Figure \ref{fig:3edge}, 
           the following shapes may be distinguished:
           an irregular shape inside the one formed 
           by a Coarse face and a Fine 2-cluster.
    }
  \label{fig:edgeboundary}
\end{figure}

%% file: simpleshapes.tex
The interpolation is performed on the shapes 
a resolution 3-edge decomposes into.
In practice, the problem is solved in the following order:
for a known decomposition pattern 
we calculate the interpolation weights for each shape involved,
starting with simpler ones.
The algorithm quits as soon as all the interpolation weights 
are positive and less than one, 
i.e. the point is inside the given shape.

The simplest shape is a tetrahedron,
for which there is only one second order accurate interpolation scheme.
It is encountered in many possible configurations,
certain central hexahedra decompose into a set of tetrahedra (see panels
\subref{fig:3edge:22},    \subref{fig:3edge:23}, \subref{fig:3edge:32},
\subref{fig:3edge:34} and \subref{fig:3edge:42} in Figure \ref{fig:3edge}).

Another simple shape is a triangular wedge prism
(see panels \subref{fig:3edge:21} in Figure \ref{fig:3edge}
and \subref{fig:edgeboundary:2C1}, \subref{fig:edgeboundary:2C2} 
in Figure \ref{fig:edgeboundary})
or a triangular prism 
(see panel \subref{fig:edgeboundary:1C} in Figure \ref{fig:edgeboundary}).
Interpolation approach is similar to that for a resolution 2-edge.
Triangular interpolation is used in the plane 
perpendicular to wedge's side edges,
while values in intersection points are obtained using linear interpolation
along these edges.

The next shape is a trapezoidal 
(see panels \subref{fig:3edge:14}, \subref{fig:3edge:24}, \subref{fig:3edge:31} 
and \subref{fig:3edge:41} in Figure \ref{fig:3edge})
or a rectangular pyramid
(see panel \subref{fig:edgeboundary:1C} in Figure \ref{fig:edgeboundary}).
Interpolation is effectively a linear interpolation 
performed on an apex and projection
of a given point along line {\it apex-point} onto a base.
The value in a projection point is obtained using either bilinear interpolation
if a given point projects onto central part of a base,
or a triangular interpolation if it projects closer to trapezoid's legs.

\begin{figure}[!h]
  \centering
  \begin{tabular}{ccc}
    \includegraphics[height=3.0cm]{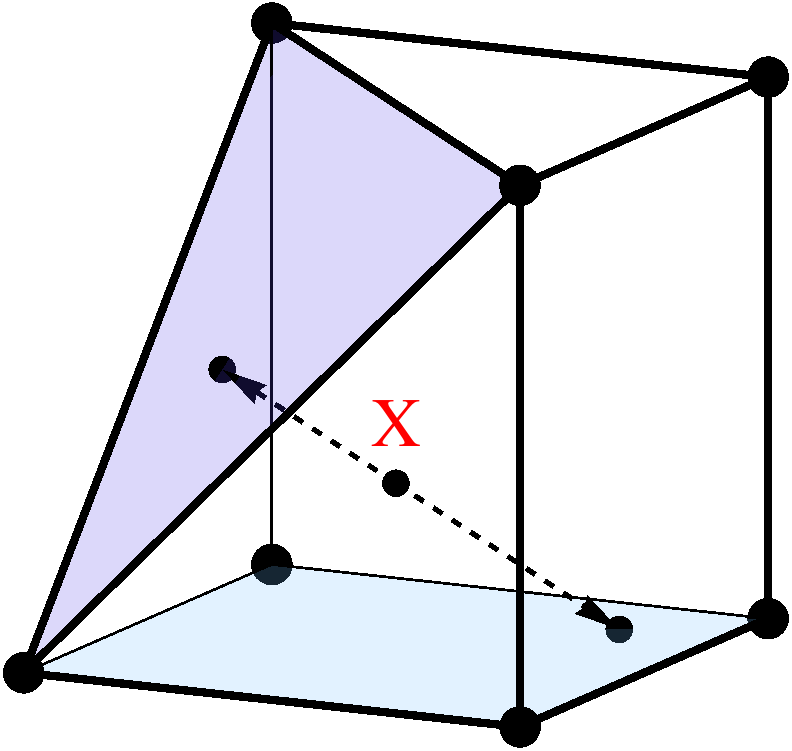} &
    \includegraphics[height=3.0cm]{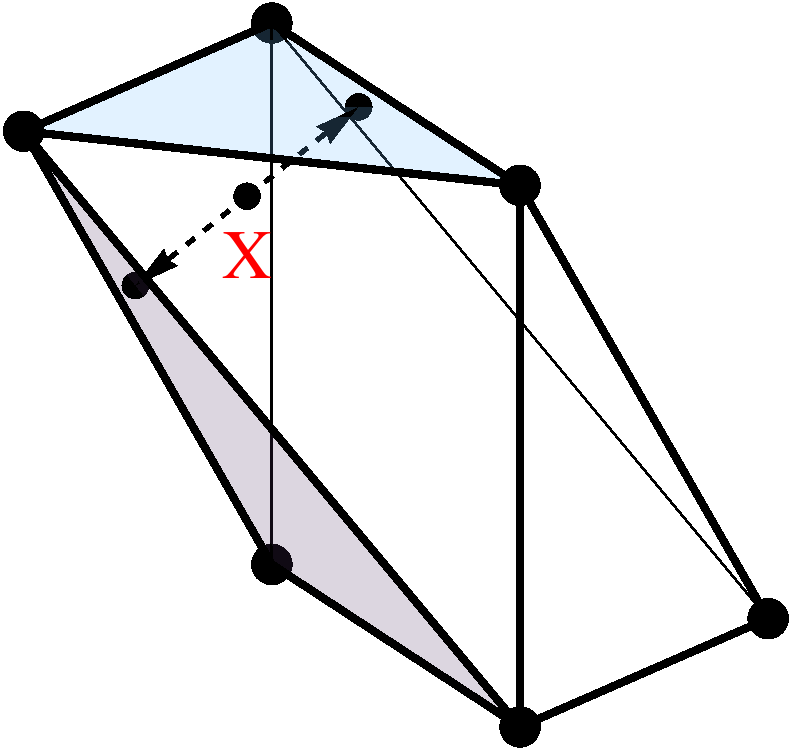} &
    \includegraphics[height=3.0cm]{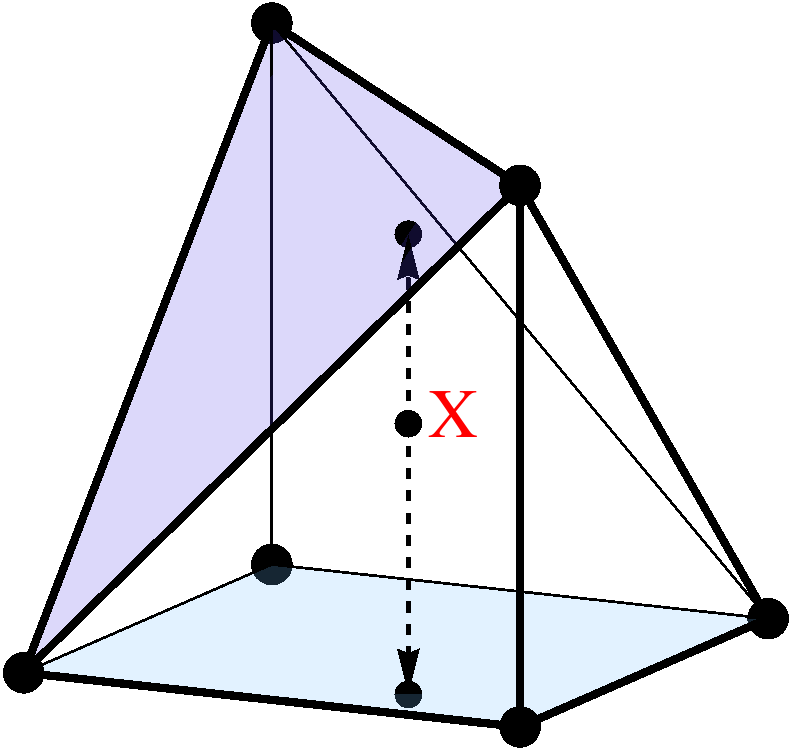}
  \end{tabular}
    \caption{An example of interpolation procedure using a ray tracing.
             Direction of the ray (dashed line) is chosen to be 
             the longest diagonal
             of the central hexahedron
             for the irregular shapes in the tessellations
             \subref{fig:3edge:11}, 
             \subref{fig:3edge:12} in Figure \ref{fig:3edge}
             and a perpendicular to a 4-cluster 
             for all the other irregular shapes
             (see panels \subref{fig:3edge:11} in Figure \ref{fig:3edge}
             and \subref{fig:edgeboundary:2C2} 
             and \subref{fig:edgeboundary:4C} 
             in Figure \ref{fig:edgeboundary}).
             Shaded faces are those intersected by the ray, 
             interpolation performed on them is either triangular, or bilinear.
             Linear interpolation is used along the ray
             to obtain a value in the point {\bf X}.
            }
  \label{fig:3DRay}
\end{figure}

For the remaining shapes the approach is the following.
The value in a given point {\bf X} is obtained via
linear interpolation on points, 
resulting from 
intersection of a ray of a certain direction going through {\bf X}
with faces of a shape.
On these faces either triangular, or bilinear interpolation is used.
Direction of a ray is chosen to be the longest diagonal 
of the central hexahedron
for the irregular shapes in the tessellations \subref{fig:3edge:11} and 
\subref{fig:3edge:12} in Figure \ref{fig:3edge}
and perpendicular to the plane of a 4-cluster 
for all the other irregular shapes
(see irregular shapes \subref{fig:edgeboundary:2C2} 
and \subref{fig:edgeboundary:4C} in Figure \ref{fig:edgeboundary}).
Examples of such interpolation on irregular shapes 
is shown in Figure \ref{fig:3DRay}.

%% file: Test.tex
To verify order and continuity of the algorithm the following test is used.

We start from a $2\times 2$ base grid in 2D or 
a $2\times 2\times 2$ base grid in 3D, 
i.e. $2^N$ grid blocks. 
Then an arbitrary subset of the $2^N$ grid blocks is refined,
giving $2^{2^N}-1$ possible AMR configurations (15 in 2D and 255 in 3D).
The order of accuracy and the continuity are checked for each configuration.


The approximation order is verified as follows.
When applied to linear functions of coordinates, 
the second-order algorithm should reproduce the function exactly,
particularly, the coordinates of an arbitrary point on the grid
are themselves linear functions
($x(x,y,z)=x$ etc.),
therefore, the coordinates of the stencil vertices,
once summed up with interpolation weights for arbitrary point {\bf X},
yield exact coordinates of this point.

Continuity of the algorithm is verified by checking the capability 
to interpolate a non-linear function of coordinates with no jump
in the interpolated values.
To each cell a random value in the range $[0.25,0.75]$ in the Finer blocks
and in the range $[0,1]$ in the Coarser blocks is assigned.
In this way the spatial gradient along direction $i$ of thus sampled function
is bounded by the value of $1/\Delta x_i^{(C)}$.
Two random points are generated, each is displaced in respect to the other 
by at most $0.01$ along each axis.
Then, for a continuous interpolation procedure,
the difference between interpolated values is 
bounded by sufficiently small value.
For the set of parameters chosen, this value is $\approx 0.01N$.

The algorithm is tested as a part of nightly tests of SWMF,
during each test 20000 random points are
generated for both 2D and 3D cases.

%% file: Conclusion.tex
We have presented a generalization of a bilinear/trilinear interpolation
to block-adaptive grids 
that is continuous and of second order accuracy.
The approach employs sequential inheritance of interpolation patterns
from lower to higher dimensions,
which makes it effective and allows us to avoid excessive logical branching
in the implementation of the algorithm.
Time performance is comparable to that of a basic bilinear/trilinear 
interpolation
applied to a uniform grid.
Thus, the algorithm is well suited to be used for interpolation 
without degrading the simulation performance.
It is applicable to the problems of data visualization as well,
such as generation of crack free isosurfaces or 
extracting smoothly varying data on cut planes and other surfaces.